\documentclass[12pt]{iopart}

\usepackage{lineno,upgreek}
\usepackage[dvipsnames]{xcolor}
\usepackage{graphicx}
\usepackage{hyperref}
\usepackage[font=footnotesize,labelfont=bf]{caption}
\usepackage{siunitx}
\usepackage{subcaption}

\expandafter\let\csname equation*\endcsname=\relax
\expandafter\let\csname endequation*\endcsname=\relax
\usepackage{amsmath}

\newcommand{\figref}[1]{Figure~\ref{#1}}

\begin{document}

\title[ETpathfinder sensitivity study]{ETpathfinder: a cryogenic testbed for interferometric gravitational-wave detectors}

\author{A. Utina$^{1,2}$, A. Amato$^{1,2}$, J. Arends$^{3}$, C. Arina$^{4}$, M. de Baar$^{5}$, M. Baars$^2$, P. Baer$^6$, N. van Bakel$^2$, W. Beaumont$^7$, A. Bertolini$^2$
% ...
}

\ead{a.utina@maastrichtuniversity.nl}
\address{$^1$ Maastricht University, Department of Gravitational Waves and Fundamental Physics, 6200 MD Maastricht, Netherlands}
\address{$^2$ Nikhef, Science Park 105, 1098 XG Amsterdam, Netherlands}
\address{$^3$ VU Amsterdam, Department of Physics and Astronomy, Vrije Universiteit Amsterdam, De Boelelaan 1085, NL-1081 HV Amsterdam, Netherlands
}
\address{$^4$ UC Louvain (Center for Cosmology, Particle Physics and Phenomenology, 2 Chemin de Cyclotron, 1348 Louvain-la-Neuve, Belgium)}
\address{$^5$ Eindhoven University of Technology, 5612 AZ Eindhoven, Netherlands}
\address{$^6$ Fraunhofer ILT - Institute for Laser Technology, Steinbachstr. 15, 52074 Aachen, Germany
}
\address{$^7$ Universiteit Antwerpen, Prinsstraat 13, 2000 Antwerpen, Belgium}
% ...

\vspace{10pt}
\begin{indented}
\item[]June 2022
\end{indented}

\begin{abstract}
The third-generation of gravitational wave observatories, such as the Einstein Telescope (ET) and Cosmic Explorer (CE), aim for an improvement in sensitivity of at least a factor of ten over a wide frequency range compared to the current advanced detectors. In order to inform the design of the third-generation detectors and to develop and qualify their subsystems, dedicated test facilities are required. ETpathfinder prototype uses full interferometer configurations and aims to provide a high sensitivity facility in a similar environment as ET. Along with the interferometry at \SI{1550}{nm} and silicon test masses, ETpathfinder will focus on cryogenic technologies, lasers and optics at \SI{2090}{nm} and advanced quantum-noise reduction schemes. This paper analyses the underpinning noise contributions and combines them into full noise budgets of the two initially targeted configurations: 1) operating with \SI{1550}{nm} laser light and at a temperature of \SI{18}{K} and 2) operating at \SI{2090}{nm} wavelength and a temperature of \SI{123}{K}.

\end{abstract}

\newpage
\section{Introduction}
 The most recent science run of the Advanced LIGO~\cite{LIGO} and Advanced Virgo~\cite{VIRGO} detectors revealed a large collection of gravitational-wave (GW) observations, bringing the total detection number up to 90 candidates~\cite{GWTC-3b} and the detection rate to about one source per week. These detectors have had a far reaching impact on Astronomy, Physics \& Gravitation. Two particular examples are: the discovery of an intermediate mass black hole, GW190521, with a progenitor situated in the pulsation pair instability gap \cite{heavy} and the discovery of GW170817~\cite{PhysRevLett.119.161101}, the first detection of a binary neutron-star merger which provided a series of insights into cosmology~\cite{Hubble}, astrophysics~\cite{Mooley2018} and nuclear physics~\cite{Pian2017}.
 
 The third-generation (3G) of ground-based gravitational-waves detectors such as the European Einstein Telescope~\cite{ET_old} and the US-based Cosmic Explorer~\cite{CE} are expected to generate a massive catalog of these binary mergers. An observatory such as ET is expected to produce a detection rate of hundreds of neutron-star coalescences and thousands of binary black-holes per week~\cite{Regimbau2012MockDC, Regimbau2014SecondET} and to explore a wider region of the parameter space that can shed light on key issues of fundamental physics, cosmology and astrophysics~\cite{sbook}.

The ET observatory aims to achieve at least an order of magnitude better sensitivity broadband compared to the current detectors and to enlarge the detection band to frequencies as low as a few Hz. Its astrophysical reach will be up to a redshift of 20 for non-spinning binaries with equal masses and a merger bound in the interval of $(20-100)$ M$\textsubscript{\(\odot\)}$~\cite{Hall_2019}. This means that at design sensitivity, ET can be sensitive to signals coming from the dark ages of the Universe, probing the birth of the first stars.

ET aims to achieve this strain sensitivity by measuring distances over a longer baseline of \SI{10}{km}. In contrast to the first and second generation of GW antennas, ET will be based on three nested detectors arranged into a triangular configuration. This will remove the blind spots in the antenna pattern and it will give equal sensitivities to both polarisation tensors of the GW signal. Each individual detector will be formed by two separately optimised interferometers for low and high signal frequencies, with medium bandwidths, forming the so-called xylophone configuration~\cite{Hild_2009}. The sum of the best individual sensitivities in the combined bandwidth forms the full ET sensitivity curve. The high-frequency detector will operate at room temperature with a high-power laser and fused-silica interferometer mirrors, called test masses, while the low-frequency detector will operate at cryogenic temperatures with lower input power impinging on the cryogenic optics made out of silicon. The entire observatory will be built underground to reduce the impact of the gravity gradient noise~\cite{Hild_2011}.

At cryogenic temperatures, the thermal noise of the test masses is reduced according to the equipartition theorem. Moreover, the broadband thermal noise can be concentrated into well-defined resonant modes for materials that have a high mechanical Q-factor~\cite{PhysRev.83.34}. The currently used fused-silica material decreases in Q factor with temperature and is not a suitable material for cryogenic detectors~\cite{Travasso2007}. On the contrary, crystalline silicon shows a higher mechanical Q factor that even slightly increases with reducing temperature~\cite{McGuigan1978}. However, silicon strongly absorbs the currently used laser wavelength of \SI{1064}{nm} and thus a change of the laser light to a longer wavelength is desirable.

This incongruity is resolved by implementing the telecommunication spectrum wavelengths at \SI{1550}{nm} and a choice of thulium- and holmium-based laser sources in the \SI{1800}{nm} to \SI{2100}{nm} range. The \SI{1550}{nm} laser wavelength is the baseline choice for the cryogenic Einstein Telescope~\cite{Hild_2011} and together with silicon test masses they constitute the building blocks of the future detectors~\cite{Schnabel_2010}. At this wavelength, the substrate absorption was measured to be as low as a few ppm/cm \cite{Degallaix:13}. Additionally, squeezed states of light, which are now used in GW detectors to reduce the impact of quantum noise have been produced at \SI{1550}{nm} with a noise-suppression factor greater than \SI{12}{dB}~\cite{Mehmet:11}. Nonetheless, further increasing the laser wavelength to \SI{2090}{nm} can bring a substantial improvement in the coating thermal noise due to the lower absorption in the highly reflective mirror coatings~\cite{PhysRevLett.120.263602} and in the reduction of the wide-angle scattering power. Advanced future detectors, such as LIGO Voyager~\cite{Adhikari_2020} or CE will have this laser wavelength as their starting choice.

ETpathfinder is a cryogenic research and development lab which consists of twenty research institutions and is funded by a consortium of financial partners\footnote{For a list of the financial partners, see Section~\ref{sec:funds}}. The facility will integrate the core technologies presented above in a \SI{10}{m} scale interferometer and characterise their compatibility with cryogenic interferometric subsystems. The compatibility is strongly linked to the challenges that need to be overcome in order to achieve sensitive interferometric operations at cryogenic temperatures. As an example, there is a clear disparity between having the test mass as isolated from the environment as possible, but in the same time having it indirectly connected to the conduction cooling system to reach the low temperatures required. The potential seismic shortcut of the cooling system needs to be resolved, with the conduction cooling being performed during operation in a controlled manner by reducing its impact onto the fundamental noise limited sensitivity of the interferometer.

In comparison with the cryogenic infrastructures for gravitational-wave detection, such as the Japanese KAGRA detector~\cite{K1,K3} and the planned cryogenic upgrade LIGO Voyager~\cite{Adhikari_2020}, ETpathfinder has close resemblances. KAGRA operations are designed for low cryogenic temperatures, such as \SI{20}{K}, integrating conduction cooling via low-stiffness heat-links, similar to ETpathfinder. However, KAGRA runs with the standard \SI{1064}{nm} laser wavelength with sapphire test masses and suspension fibres. The LIGO Voyager is designed to run at around \SI{2000}{nm} laser wavelength, with silicon test masses kept at about \SI{123}{K}, configuration similar to ETpathfinder-B.

This paper explains the design displacement sensitivity towards the ETpathfinder at \SI{18}{K} mirror temperature and \SI{1550}{nm} laser light as well as \SI{123}{K} for \SI{2090}{nm} wavelength. In Section~\ref{sec:facility} the infrastructure and optical layout of the facility is presented. This is followed by the main section, Section~\ref{sec:Sec3}, where each noise source is described and its constituent parameters shown. The paper ends with a short discussion about the main findings presented in Section~\ref{sec:Conclusions}.

\section{The ETpathfinder facility}
\label{sec:facility}
\subsection{Infrastructure}
\label{sec:Infrastructure}
ETpathfinder is being built in a former warehouse in the Randwyck area of Maastricht, the Netherlands. The building's flooring has been replaced with a \SI{45}{cm} thick concrete slab resting on around 170 pillars that reach about \SI{5}{m} into the ground and is decoupled from the building walls in order to reduce tilting of the floor. On top of this concrete slab rests the purpose-built ISO 7 and 8 cleanroom with dedicated air handling and climate control and a floor space of \SI{850}{\square\meter}. A separate \SI{64}{\square\meter} area is reserved for preparation,  cleaning and baking of parts that go into the main cleanroom and vacuum system. Large items, such as the vacuum system itself, enter through a goods reception area that can be accessed from the outside and is also reachable with a $2\times\SI{2}{tonne}$ crane that is installed in the cleanroom. Thanks to the multiple lock system, the large components can be moved into the cleanroom without opening it to the outside or loosing the over-pressure. Noisy equipment, such as vacuum pumps, are located in a separate room whose floor is decoupled from the experimental area in an effort to reduce vibrational coupling to a minimum.

\subsection{Optical Layout}
\label{sec:layout}
\begin{figure}[htbp]
    \centering
    \includegraphics[width=\linewidth]{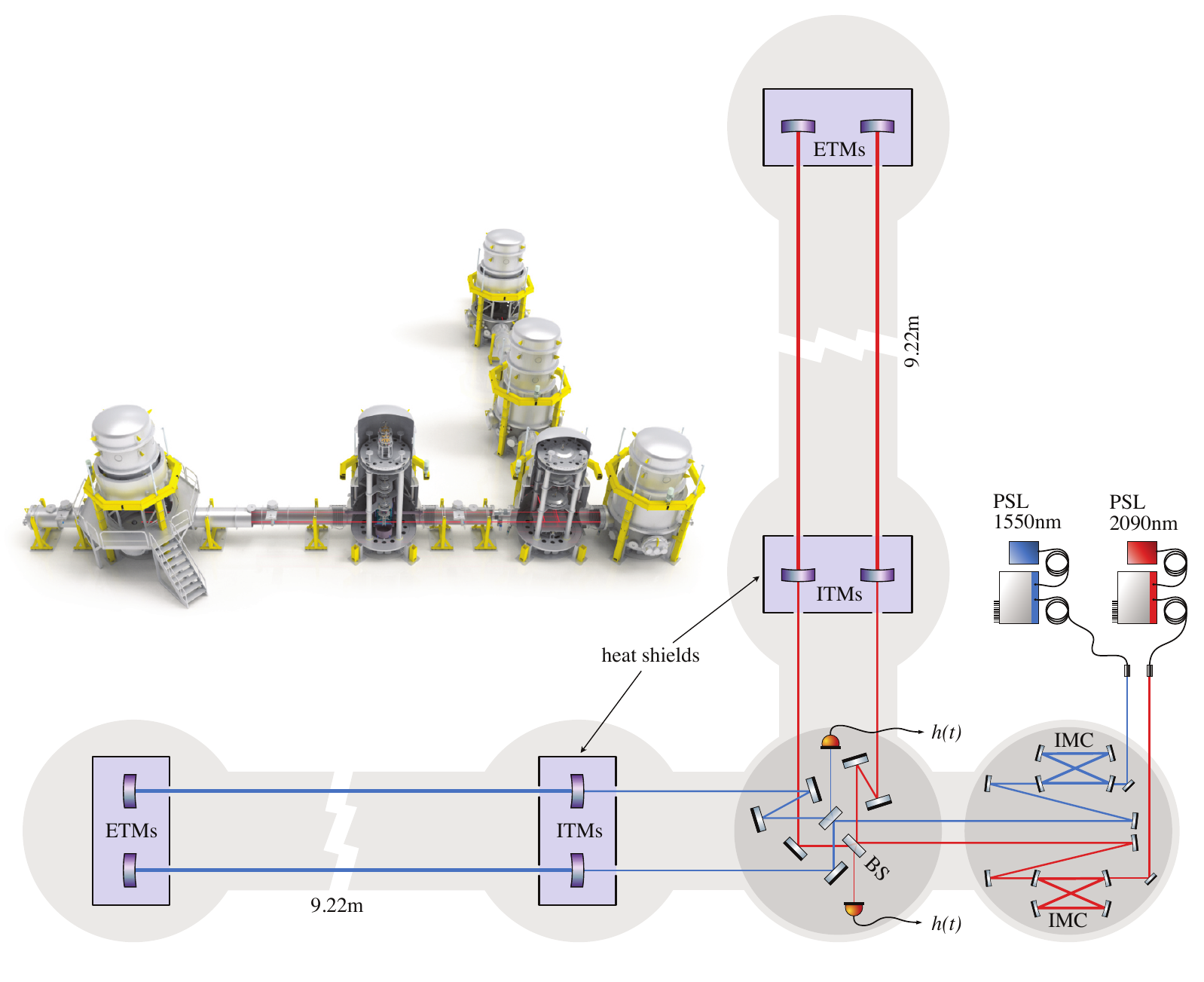}
    \caption{Simplified layout of ETpathfinder. Each of the two interferometers, here distinguished by the blue (\SI{1550}{nm}) beams and red (\SI{2090}{nm}) beams, occupies one of the two arms of the L-shaped vacuum system, with shared beam-splitter bench and input-optics bench. The input test-masses (ITMs) and end test-masses (ETMs) are separated by a distance of \SI{9.22}{m} and situated into cryostats (light blue boxes). Light is provided by two pre-stabilized laser sources (PSLs), outputting around \SI{1}{W} each. Inside the vacuum system, the laser beams are further stabilized and cleaned by input mode-cleaners (IMCs), before mode-matching telescopes guide the beams towards the beam-splitter and arm resonators. The interferometer outputs are kept close to a dark fringe and are detected by photodiodes. Their signal is proportional to the difference in arm-length (DARM), providing the main science signal $h(t)$. The inset on the left shows a cut-away rendering of the actual setup.}
    \label{fig:ol}
\end{figure}

ETpathfinder consists of two Michelson interferometers with Fabry-Perot
resonators in the arms. One interferometer will operate at \SI{1550}{nm} (here
labelled ETpathfinder-A) and the second one at \SI{2090}{nm} laser wavelength
(ETpathfinder-B). Both interferometers are co-located in the same vacuum
envelope, which is schematically presented in \figref{fig:ol}. In total there
are six vacuum towers, of which two, the input-optics tower and the
beam-splitter tower, contain suspended optical benches, while the other four
towers contain the individually suspended input and end test mass mirrors (ITMs
and ETMs, respectively). The test masses are placed inside cryostats formed by multiple double-layered thermal shields. The interferometer arms are folded to run in parallel,
this way the two interferometers can be operated at different temperatures with both test masses inside the same cryostat. Additionally, a single Fabry-Perot Michelson interferometer can be operated with heavier test masses, each per cryostat.

The light sources for the interferometers are pre-stabilized lasers (PSLs)
located on an optical table outside of the vacuum system. They provide around
\SI{1}{W} of continuous-wave input light in the TEM\textsubscript{00} fundamental
mode. Control loops reduce the free-running laser noise by several orders of
magnitude for both frequency and amplitude noises in the frequency band of
interest, i.e.\ between \SI{10}{Hz} and \SI{10}{kHz}. A detailed description of
the \SI{1550}{nm} system is given in~\cite{Meylahn22}.

Inside the vacuum system, the laser beams are first guided to input mode-cleaners, four-mirror ring resonators which provide a stable beam reference, reducing beam-pointing noise and serving as a quiet frequency reference for further stabilization of the laser noise. Mode-matching telescopes adjust the beam shape to match the eigenmode of the arm cavities, then send the beam to the beam-splitter bench. The central interferometer optics, including the beam-splitters, are mounted on
in triple-suspension cages closely modelled after Advanced LIGO's HRTS design~\cite{HRTS}. A total of three folding mirrors for each interferometer help to guide the beam into the arm cavities.

The ETM and ITM mirrors are made of silicon, with a diameter of \SI{150}{mm} and a thickness of \SI{80}{mm}, and are spaced \SI{9.22}{m} apart. Taking the density of crystalline silicon to be $\SI{2329}{kg\per m^{3}}$, a mirror mass of \SI{3.29}{kg} results. The highly reflective surface of the mirrors has a radius of curvature of \SI{14.5}{m}, which results in a symmetric cavity with a beam waist size of \SI{1.8}{mm} for the \SI{1550}{nm} interferometer, giving a beam radius of \SI{2.2}{mm} on the mirrors. Furthermore, the beam dimensions will scale with $\sqrt{\frac{2090}{1550}}$ for the \SI{2090}{nm} interferometer since this has the same radii of curvature for the test masses and the same Rayleigh range as the \SI{1550}{nm} one.  To counteract the beam divergence due to the high refractive index of 3.5 of the test masses in both arms, the anti-reflective coated back surfaces have a curvature of radius \SI{9.0}{m}, leading to a collimated beam leaving the arm cavities.

\section{The fundamental noise sources of ETpathfinder}
\label{sec:Sec3}
\begin{figure}[ht]
     \centering
     \begin{subfigure}[b]{1\textwidth}
         \centering
         \includegraphics[width=.7\textwidth]{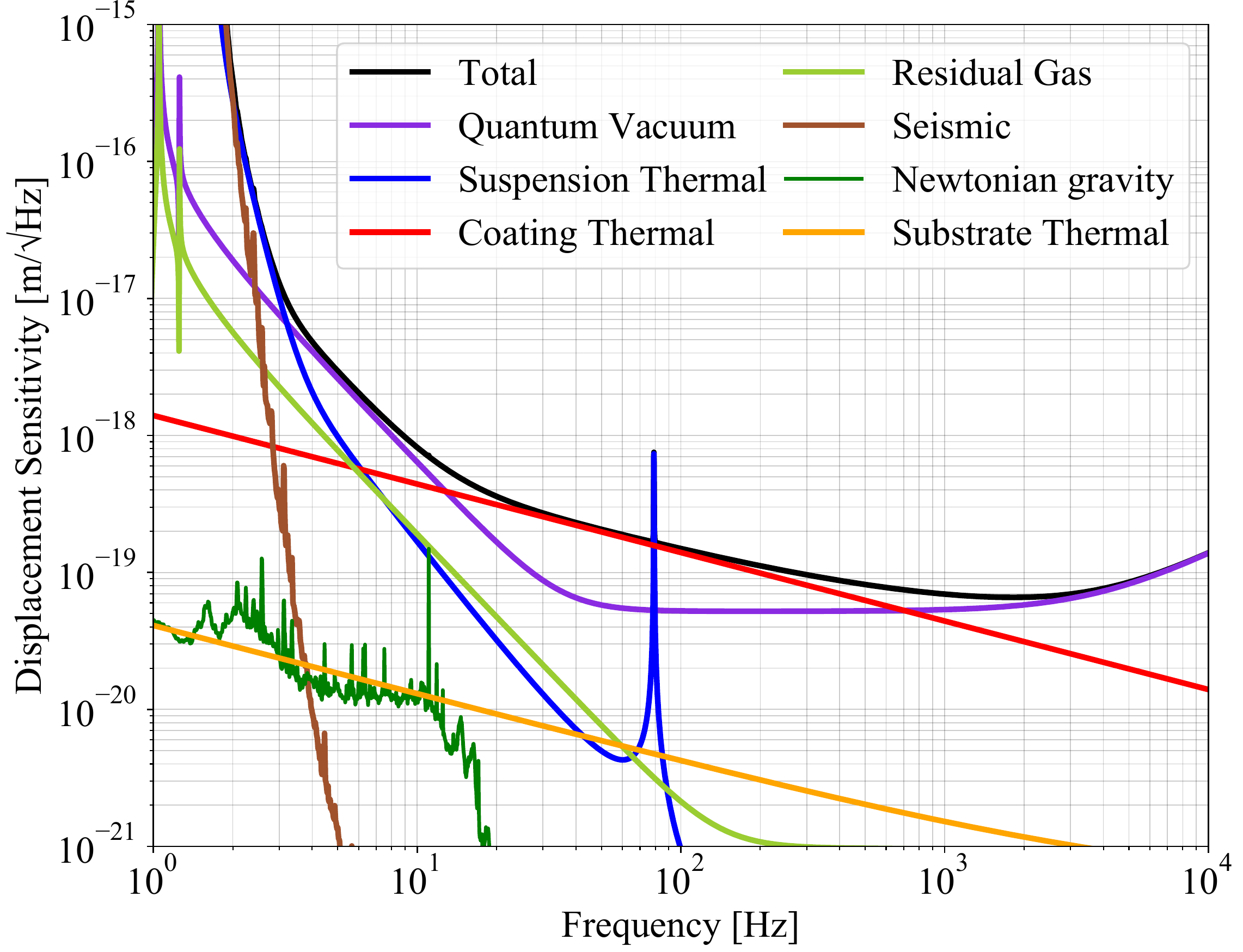}
         \caption{}
     \end{subfigure}
     
     \hfill
     \begin{subfigure}[b]{1\textwidth}
         \centering
         \includegraphics[width=.7\textwidth]{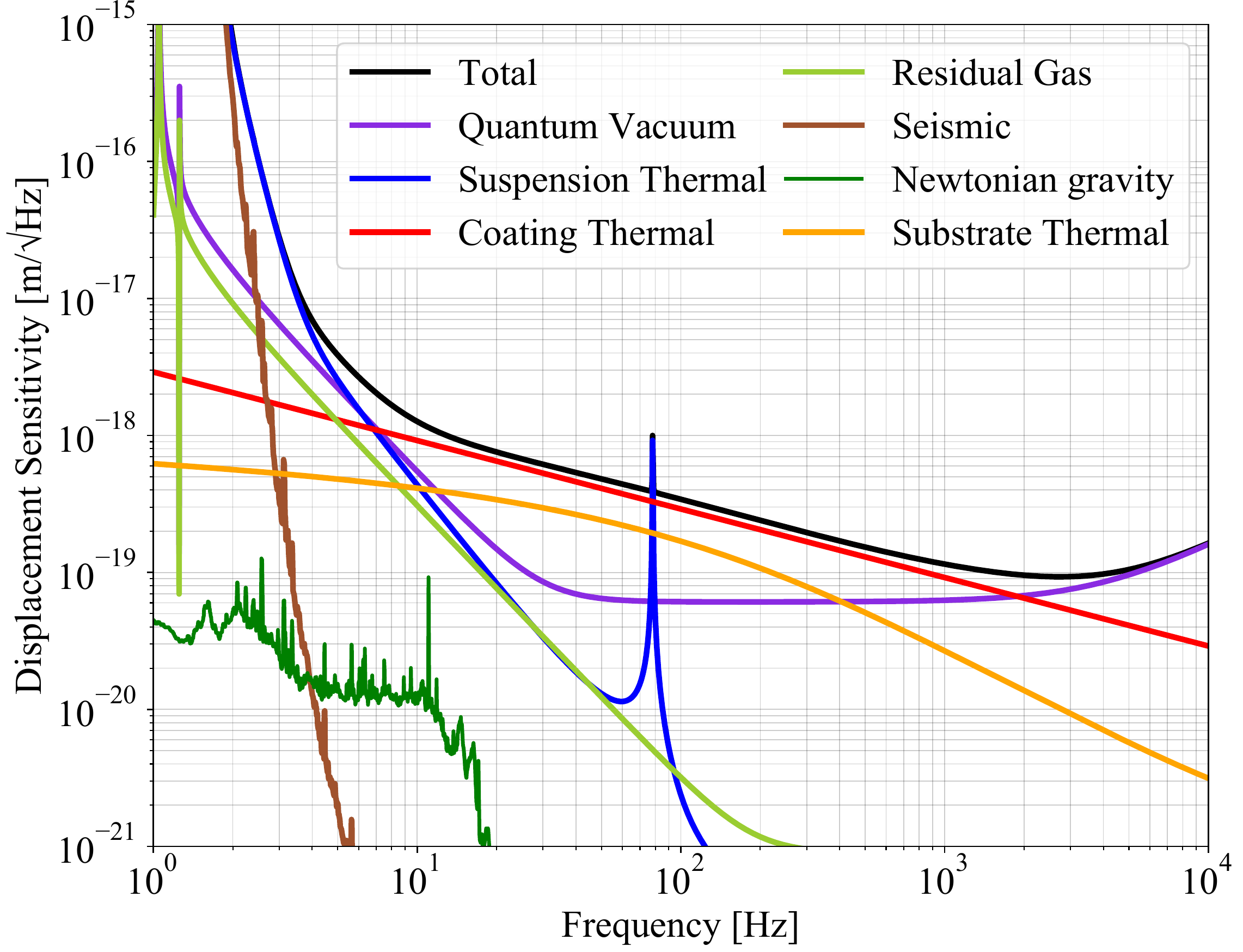}
         \caption{}
    \hfill
     \end{subfigure}
        \caption{Projection of displacement sensitivity for the ETpathfinder operating at \SI{1550}{nm} laser light and \SI{18}{K} (a) and at \SI{2090}{nm} and radiatively cooled down at \SI{123}{K} (b).}
        \label{fig:ETpfAB}
\end{figure}

As a prototyping platform for third-generation gravitational-wave detectors,  ETpathfinder aims to reach a displacement sensitivity that allows for the testing of novel techniques for the cool-down and control of cryogenic interferometers at realistic and meaningful noise levels. The impact of these techniques on the fundamental sensitivity characterizes their compatibility level. Therefore, we aim to reduce all fundamental noise sources to a level below $\SI{1e-18}{m\per\sqrt{Hz}}$ at \SI{10}{Hz} for the interferometer operating at \SI{1550}{nm} and at \SI{18}{K}. While the \SI{2090}{nm} interferometer during operation 
using cryogenic silicon suspensions will be close to reaching the target, as presented in Figure~\ref{fig:ETpfAB}, the \SI{18}{K} interferometer at \SI{1550}{nm} will be able to reach the low-noise sensitivity broadband. Reaching this configuration in a single step is very challenging, therefore we first aim for an initial running phase with simplified mirror suspensions and a higher operating temperature of \SI{123}{K} that allows for radiative instead of conductive cooling. For this configuration, the expected design sensitivity is shown in~\figref{fig:ETpfA_light}, reaching the $\SI{1e-18}{m\per\sqrt{Hz}}$ target from frequencies of \SI{20}{Hz} onwards.  In the following subsections, each individual component of the noise budget is discussed elaborately.

\subsection{Quantum noise}
Quantum noise originates from fundamental fluctuations of the electromagnetic vacuum which is the manifestation of the Heisenberg uncertainty principle~\cite{PhysRevD.23.1693}. Its impact on the sensitivity of any interferometer consists of two parts: Quantum \textit{shot noise} arises from the quantization of light that comes at the photo-detector, with a fluctuation in the observed power which scales with the number of photons arriving at the detector dark port. The quantum shot noise-limited sensitivity scales down with laser power interacting with the test masses of the interferometer. Shot-noise phase fluctuations are frequency-independent and therefore have the greatest impact on the sensitivity of the interferometer at high frequencies, where signal response wanes due to finite bandwidth of the meter.

The second component of quantum noise known as \textit{quantum radiation-pressure} or \textit{quantum back-action} noise, is driven by the amplitude fluctuations of vacuum entering the interferometer at each optical loss in the components. When beating with the carrier light in the arms, these fluctuations give rise to random radiation pressure force on the mirrors, resulting in the random displacement of the mirrors and of the arm length change that mimics the signal. It scales up naturally with laser power and is most prominent at the low frequencies, where mirror displacement is larger for a given force.

The fact that quantum shot noise and then quantum radiation-pressure noise have an inverse dependence on power and that the underlying fluctuations of phase and amplitude are uncorrelated, gives rise to a so-called standard quantum limit (SQL)~\cite{Braginsky1968, Danilishin2012} on continuous high-precision interferometric measurements.

The baseline quantum-noise curve in shown in Figure~\ref{fig:ETpfAB}. A laser input power to \SI{1}{W} at the beam-splitter is assumed. An arm-cavity finesse of $\approx 2050$ is used with the circulating power
impinging on to the \SI{3.29}{kg} test masses. The radiation-pressure noise is proportional to the intra-cavity power which scales with the cavity finesse, thus limits the sensitivity from \SI{4} to \SI{10}{Hz} for ETpathfinder-A. The shot noise-limited sensitivity is enhanced by the increased circulating power below the arm-cavity pole at about \SI{3.9}{kHz}. The estimation also assumes an average optical loss of 50\,ppm per mirror.


\subsection{Coating Brownian noise}
Coating Brownian noise, together with other coating and substrate noises, are thermal displacement noises. These are best described by the Fluctuation Dissipation (FD) Theorem. This demonstrates a relationship between the amount of fluctuation of the test mass surface and the dissipation in the system~\cite{PhysRev.83.34}:

\begin{equation}
    S(f)=\frac{k_{B} T}{\pi^{2} f^{2}} {Re}[Z(f)^{-1}],
\label{eq:FDT}
\end{equation}
where $S(f)$ is the power spectral density of the fluctuations in any degree of freedom, $k_{B}$ is the Boltzmann constant, $T$ is the temperature of the mass and $Z(f)$ is the complex impedance of the system, or the inverse of the mechanical admittance.

The coating Brownian noise arises from the thermal motions of the individual particles in the coating material. A simplified form of the spectral density fluctuation, $S_{x}(f)$, is given by \cite{Harry_2002}:

\begin{equation}
S_{x}(f)=\frac{2 k_{B} T}{\pi^{2} f} \frac{d}{w^{2}} \phi \left(\frac{Y_{\rm coat }}{Y_{\rm sub}^{2}}+\frac{1}{Y_{\rm coat }}\right).
\end{equation}
Here, $T$ is the coating temperature, $w$ is the beam radius at the coating, as described in Section~\ref{sec:facility}, $\phi$ is the mechanical loss angle of the coating, $d$ is the coating thickness and $Y_{\rm coat}$, $Y_{\rm sub}$ being the Young's modulus of the material coating and substrate respectively. For the coating thermal noise calculations in Figures~\ref{fig:ETpfAB} and~\ref{fig:ETpfA_light} the full model described by Hong \cite{Hong} was used.

For the noise budgets presented for ETpathfinder A and B in Figs.~\ref{fig:ETpfAB} and~\ref{fig:ETpfA_light}, a conservative thermal-noise level based on well-known materials is assumed: SiO$_2$ for the low refractive-index material and Ta$_2$O$_5$ for the high refractive-index material. The optical properties of these materials are excellent. For coatings produced by LMA\footnote{Laboratoire des Matériaux Avancés, Lyon} via ion beam sputtering, optical absorptions as low as about a quarter of a ppm ($10^{-6}$) have been shown~\cite{Degallaix2017}.

Both SiO$_2$ and Ta$_2$O$_5$ are materials which have been characterized well for use in gravitational-wave detectors. SiO$_2$ shows a very low mechanical loss of $0.46 \times 10^{-4}$ at room temperature~\cite{PhysRevD.93.012007}. Ta$_2$O$_5$ is known to have a higher mechanical loss of $2.3 \times 10^{-4}$ at room temperature~\cite{Flaminio_2010}. Both materials show an increase in mechanical loss at low temperatures. The mechanical losses at the temperatures planned for ETpathfinder are given in Table~\ref{tab:parameters}.

Using those low-temperature mechanical losses, coating thermal noise would result in the level represented by the red lines in Figs.~\ref{fig:ETpfA_light} and~\ref{fig:ETpfAB}. In both cases, the total noise would be dominated by coating thermal noise over a wide frequency range between about 40\,Hz and 2\,kHz. In this noise budget, for the ETM mirrors, 17 pairs of layers are used to achieve high reflectivity. For the ITM mirrors, lower in reflectivity, 9 pairs of layers are to be used. The layers have an optical thickness, which is the actual thickness $d$ multiplied with the refractive index $n$, of a quarter of the interferometer wavelength, except for the outermost layer pair of the ITM, which was tuned for slightly lower reflectively to result in an arm cavity finesse of $\approx 2050$.

While this conventional approach of using SiO$_2$ and Ta$_2$O$_5$ in the first stage of ETpathfinder would offer the interesting option to directly observe coating thermal noise, another attractive option is to test other materials or concepts explored within the gravitational-wave community, which have not yet been implemented into any detector. The following section will explore some of these options.

\subsubsection{Other possible coating options}

\begin{figure}[htbp]
    \centering
    \includegraphics[width=.7\textwidth]{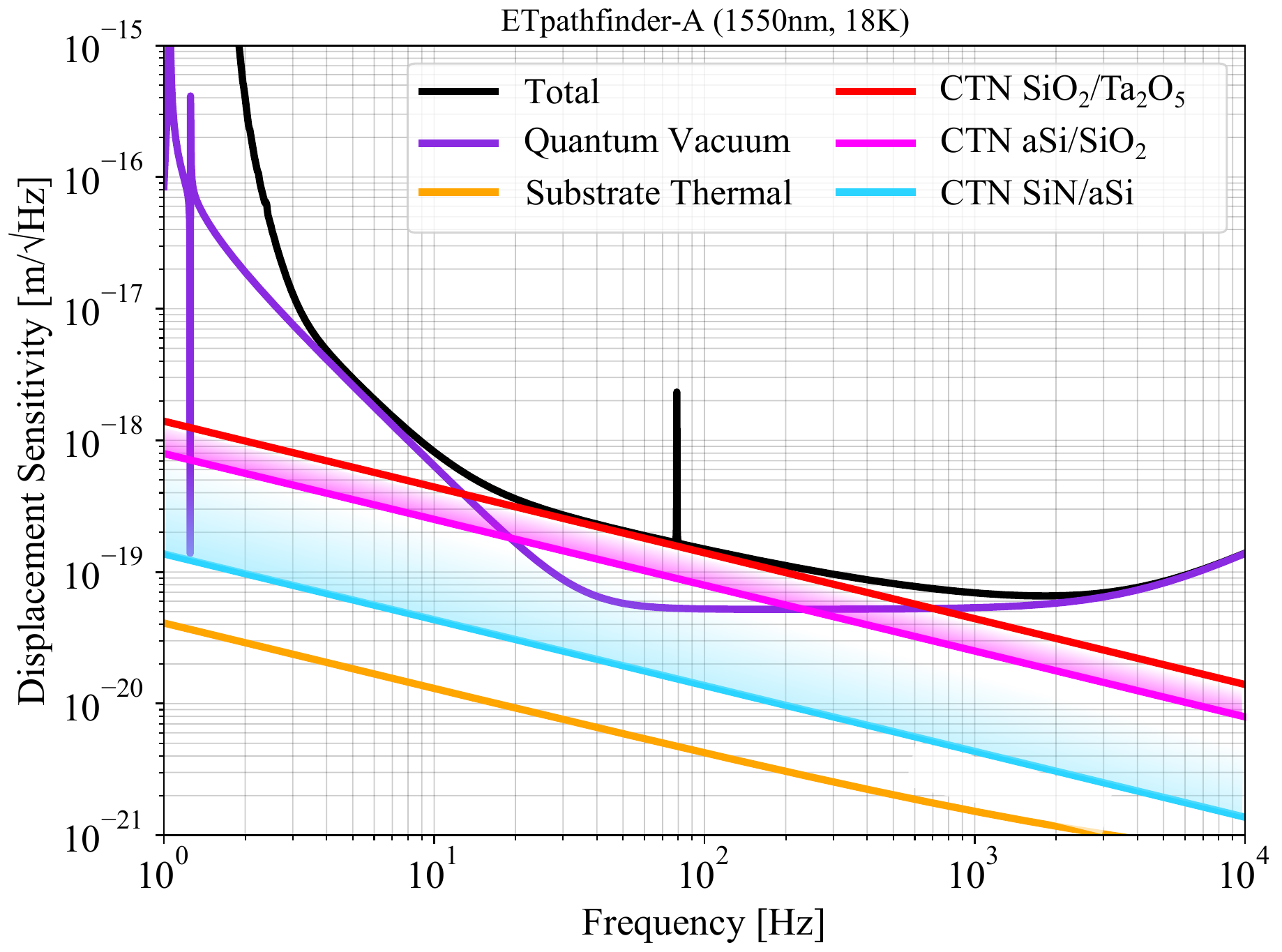}
    \caption{Modified version of Fig.~\ref{fig:ETpfAB} (a): Shown are the coating thermal noise (CTN), and the total noise and quantum noise for comparison. Substrate thermal noise is also shown as another mirror-based noise source. While the red line shows CTN for conventional SiO$_2$/Ta$_2$O$_5$ coatings, the pink line shows CTN for coatings with the same reflectivity, but using aSi instead of Ta$_2$O$_5$. The blue line shows CTN for coatings made of aSi and SiN$_x$. The shaded areas indicate the range of CTN when using multi-material designs for absorption reduction.}
    \label{fig:CTN}
\end{figure}

There are several other interesting coating options to consider for being tested in ETpathfinder:

\begin{itemize}
\item Amorphous silicon (aSi) is a material with a very low mechanical loss~\cite{Liu_1997,Birney_2018,ion_plating} and a high refractive index of $n \geq 3.5$, resulting in fewer layers being required to achieve high reflectivity when using it in a stack with SiO$_2$. Both the lower mechanical loss and the reduced stack thickness reduce coating thermal noise, resulting in the pink line shown in Fig.~\ref{fig:CTN}. Coating thermal noise of this coating is dominated by the SiO$_2$ layers.
The optical absorption poses an obstacle for the use of aSi in gravitational-wave detectors. The lowest absorption realized so far corresponds to 7.6\,ppm for a highly-reflective coating made of aSi combined with SiO$_2$ at 1550\,nm~\cite{Birney_2018}. This is very likely still too high for the use of aSi in the Einstein Telescope. While it may be a tolerable absorption level for ETpathfinder, such low absorption has only been realized on an R\&D level so far and was not yet reproducible. Commercially available aSi may still show too high absorption even for ETpathfinder~\cite{Steinlechner2016,ion_plating}.
Nevertheless, using aSi in ETpathfinder would be an interesting step forward for using this material in future gravitational-wave detectors.
\item Another interesting amorphous material is silicon nitride (SiN$_x$), which shows a low mechanical loss similar to that of aSi, but also comes with similar absorption issues~\cite{Pan_2018,liu_metcalf_wang_photiadis_2007,PhysRevD.96.022007}. The refractive index of SiN$_x$ is similar to that of Ta$_2$O$_5$, varying with deposition method and exact stoichiometry. Therefore, SiN$_x$ could serve as a high refractive-index material together with SiO$_2$ or as a low refractive-index material together with aSi. The latter would result in very low coating thermal noise, see blue line in Fig.~\ref{fig:CTN}, but even at $\approx$2\,$\upmu$m, where the absorption is lower than at 1550\,nm, it would be at a few ten ppm~\cite{ion_plating}.
\item To avoid too high absorption levels and to test another interesting concept future gravitational-wave detectors may benefit from, the implementation of multimaterial coatings can be considered for ETpathfinder: In such coatings a few layer-pairs of low-absorption materials, e.g. SiO$_2$/Ta$_2$O$_5$, are used to reflect the majority of the laser power before materials with higher absorption but low mechanical loss, such as aSi or SiN$_x$, are used further down in the coating to reduce the overall coating thermal noise~\cite{PhysRevD.91.042001,PhysRevD.91.042002,PhysRevLett.125.011102}. This option is indicated by the pink shaded and blue shaded areas. While pure aSi/SiO$_2$ or aSi/SiN$_x$ coatings would result in the pink and blue solid lines, with every pair of SiO$_2$/Ta$_2$O$_5$ to reduce the absorption, the coating thermal noise level converges towards the level of SiO$_2$/Ta$_2$O$_5$ only, shown by the red line. The exact level depends on the number of SiO$_2$/Ta$_2$O$_5$ required to achieve a certain absorption level (which in turn depends on the coating material properties and on the tolerable absorption level).
\item Another very promising coating option for gravitational-wave detectors are crystalline coatings. AlGaAs/GaAs coatings show excellent thermal noise and optical absorption~\cite{Cole2016}, however, the challenge is that crystalline coatings have to be grown on lattice-matched crystalline substrates (and afterwards transferred and bonded to a suitable mirror substrate). In case of AlGaAs/GaAs this would be a GaAs wafer, which is not available in sizes required for gravitational-wave detectors. While coatings in sufficient sizes for ETpathfinder could be realized, there is no major benefit in testing this well-characterized, but likely not sufficiently upscalable material. However, new material combinations which can be grown on larger substrates, or even directly on crystalline detector mirrors, are under investigation within the ETpathfinder project. Crystalline coatings may reach a coating thermal noise level similar to that of aSi/SiN$_x$ shown by the blue line in Fig.~\ref{fig:CTN} or even below.
\end {itemize}

\subsection{Substrate noises}
Mechanical dissipation inside the test mass material drives noise sources associated with the substrate. As these are thermal noises, they can be quantified using the FD theorem. Here, three substrate noise sources for crystalline silicon test masses are described: the ITM-thermorefractive noise, the substrate Brownian and thermoelastic noises.

Thermorefractive noise is the optical path length change from the variations in the refractive index of the test mass due to random fluctuations of temperature in the refracting material. This phase noise is transmitted through the ITMs into the cavities and will degrade the sensitivity of the differential measurement. Here, the power spectral density of the ITM thermorefractive noise, $S^{\delta z}(f)$ is estimated from the exact series expansion~\cite{BRAGINSKY2004345}:
\begin{equation}
S^{\delta z}(f) = \frac{a\beta^{2} k_{b} T^{2}}{2\pi\kappa}\times \left( e^{it} E_{1}(it) + e^{-it} E_{1}(-it)    \right).
\end{equation}
Here $\beta=\partial n / \partial T$ with $n$ being the material refractive index, $a$ is the test mass thickness, $k_{B}$ is the Boltzmann constant, $T$ is the substrate temperature and $\kappa$ is the thermal conductivity. The term $t$ here is defined as  $t= \frac{2\pi fw^{2} \rho C}{2\kappa}$. Inside this term, $w$ is the laser beam size on the substrate surface, $\rho$ is the density of the material and $C$ is the specific heat. The functions $E_{1}(x)$ are exponential integrals which equal $E_{1}(x) =  \int_{x}^{\infty} e^{-xy}y^{-1} \,dy $. For the effect on both ITMs, the noise is then projected to the differential arm-length signal as $S_{\text{DARM}}^{\delta z}(f) = S^{\delta z}(f)\times \frac{\pi}{F} $, with $F$ being the arm cavity finesse value. The parameters involved in the computation are shown in Table~\ref{tab:parameters}.

\begin{figure}[htbp]
    \centering
    \includegraphics[width=.7\textwidth]{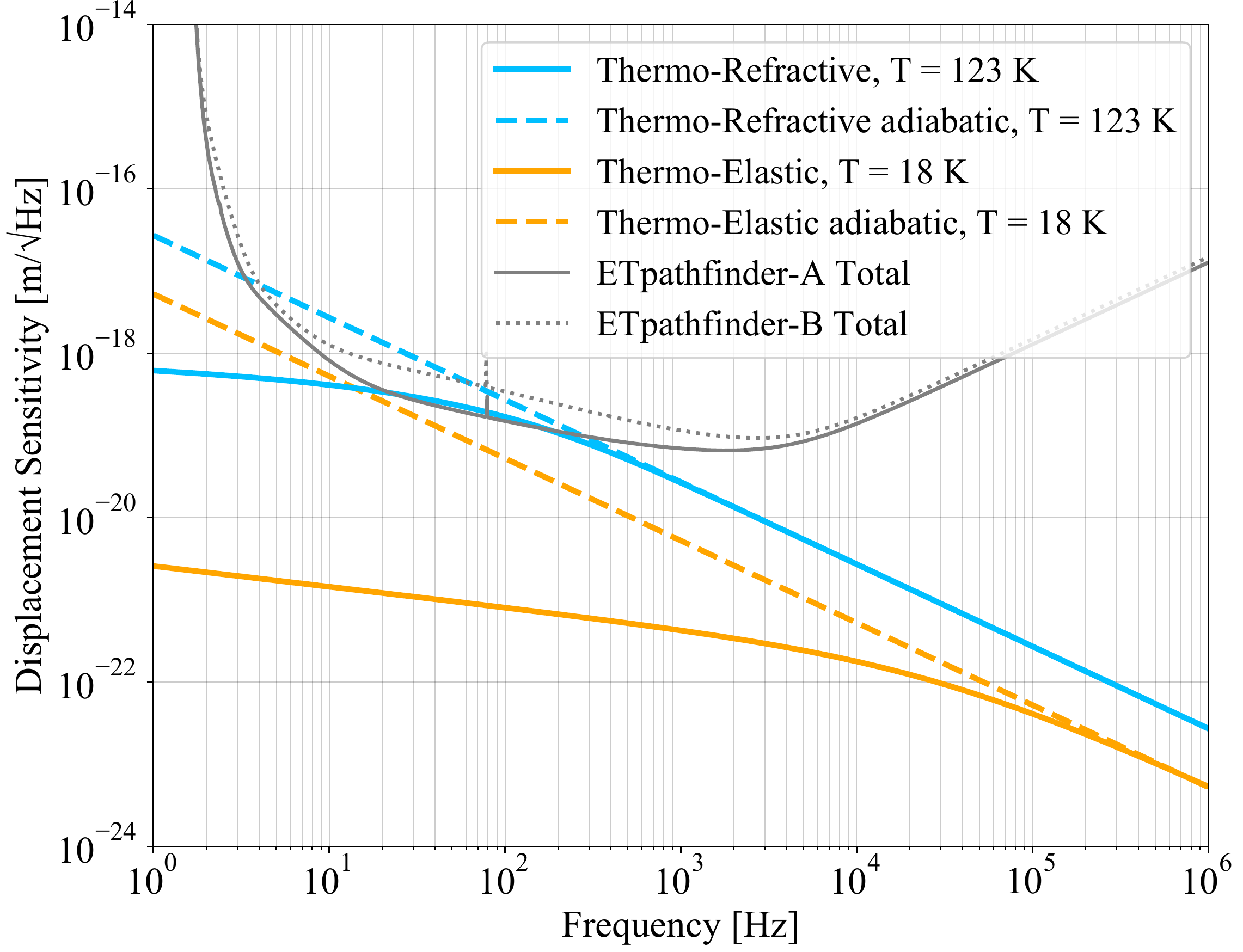}
    \caption{Displacement noises due to the ITM thermorefractive effect and the substrate thermoelastic noise when deviate from the adiabatic approximation. A substantial noise reduction is observed at frequencies below the adiabatic limit $\omega_{c}$. For silicon, $\omega_{c}$ increases with decreasing temperature as the thermal conductivity maintains high values while the specific heat of the material sharply decreases.}
    \label{fig:substrates}
\end{figure}

Furthermore, the time-averaged power dissipated into the material creates a temperature gradient which causes the material substrate to deform via the thermal expansion coefficient. Initially developed by Braginsky \textit{et al} \cite{BRAGINSKY19991}, known as BGV formulation, the treatment approximates the volume involved in the fluctuations as being of the order of the thermal diffusion length, a function of material thermal conductivity $\kappa$, its density $\rho$ and specific heat $C$ for the characteristic time $1/f$: $\sqrt{\frac{\kappa}{\rho C f}}$. This is a good approximation for room temperatures and large beam spot sizes at the mirror, but in the opposite regime, when the adiabatic approximation is not satisfied, corrections have to be implemented. For low temperatures and small beam sizes hitting the silicon mirrors of the ETpahfinder, a clear departure from the BGV approximation is observed. At \SI{18}{K} temperature, the thermal diffusion length at \SI{10}{Hz} is about two orders of magnitude higher than the beam dimension at the test mass. Moreover, at the same measurement frequency and \SI{123}{K} temperature, the diffusion length is about a factor of two higher. Thus, the full solution from \cite{Cerdonio2001} is implemented here with the spectral density of the displacement noise can be estimated as :

\begin{equation}
    S_{\mathrm{TE}}(f)=\frac{8}{\sqrt{2 \pi}} \alpha^{2}(1+\sigma)^{2} \frac{k_{B} T^{2} r_{0}}{\rho C a^{2}} J[\Omega],
\end{equation}

with $J[\Omega]$ being defined as:

\begin{equation}
    J[\Omega]=\sqrt{\frac{2}{\pi}} \int_{0}^{\infty} d u \int_{-\infty}^{\infty} d v \frac{u^{3} e^{-u^{2} / 2}}{\left(u^{2}+v^{2}\right)\left(\left(u^{2}+v^{2}\right)^{2}+\Omega^{2}\right)}.
\end{equation}
In the above equations $C$ is the specific heat of the material, $\alpha$ is the coefficient of linear expansion, $\rho$ is the density of the substrate, $r_{0}$ is the beam radius at the test mass and $a = r_{0}^{2}\cdot \omega_{c}$ with $\omega_{c}$ being the angular frequency of the adiabatic limit given by $\omega_{c} = \frac{\kappa}{\rho C r_{0}^{2}}$. Figure~\ref{fig:substrates} illustrates the low-frequency discrepancy between the adiabatic approximations and the full solutions, which makes the implementation of the complete solution strongly relevant for the parameters in our experiment.

Finally, the spectral density of the noise source associated with the Brownian fluctuations in the bulk material is estimated using an extension of Levin's direct method~\cite{Levin1998InternalTN}, developed by Bondu et al. (1998)~\cite{BONDU1998227}:

\begin{equation}
S_{x}(f)=\frac{2 k_{\mathrm{B}} T}{\sqrt{\pi^{3}} f} \frac{1-\sigma^{2}}{Y w} \phi_{\text {substrate }}(f, T).
\end{equation}
Here $\sigma$ is the Poisson ration of the material, $Y$ is the Young modulus and $\phi_{\text{substrate}}$ is the substrate loss angle. Finite test mass corrections to the thermoelastic and Brownian noises were derived by Liu and Thorne~\cite{Liu2000ThermoelasticNA} and are applied to the current calculations inside pygwinc~\cite{pygwinc}.

For test masses of dimensions presented in Section~\ref{sec:layout} and thermal dependent parameters shown in Table~\ref{tab:parameters}, the above three noises are estimated. The thermorefractive noise dominates the other substrate noises at \SI{123}{K}. This is due to the combination of a small beam radius at the substrate and the large value of the thermorefractive coefficient $\beta=\partial n / \partial T$ of $1\cdot 10^{-4}$ at \SI{123}{K} \cite{dndT}. However, at lower temperatures, because of vanishingly small linear expansion coefficient and a thermorefractive coefficient of the order of $10^{-6}$, the substrate thermal effects have negligible noise contributions to the total sensitivity curve.

\subsection{Seismic noise}
A combination of the seismic and Newtonian gravity noise dominates the noise sensitivity of the ET design, as well as current detectors at very low frequencies.  Since the beginning of the field, the seismic impact on the test masses was recognized as a major noise source \cite{Raiss} and is the main driver for the extremely advanced and delicate suspension systems seen today in gravitational-wave detectors. In this work, the noise spectral density is calculated from seismic measurements of the ambient seismic field which is propagated via the mechanical transfer function of the planned suspension system.

\subsubsection{Seismic Measurement}
\begin{figure}[htbp]
    \centering
    \includegraphics[width=1\linewidth]{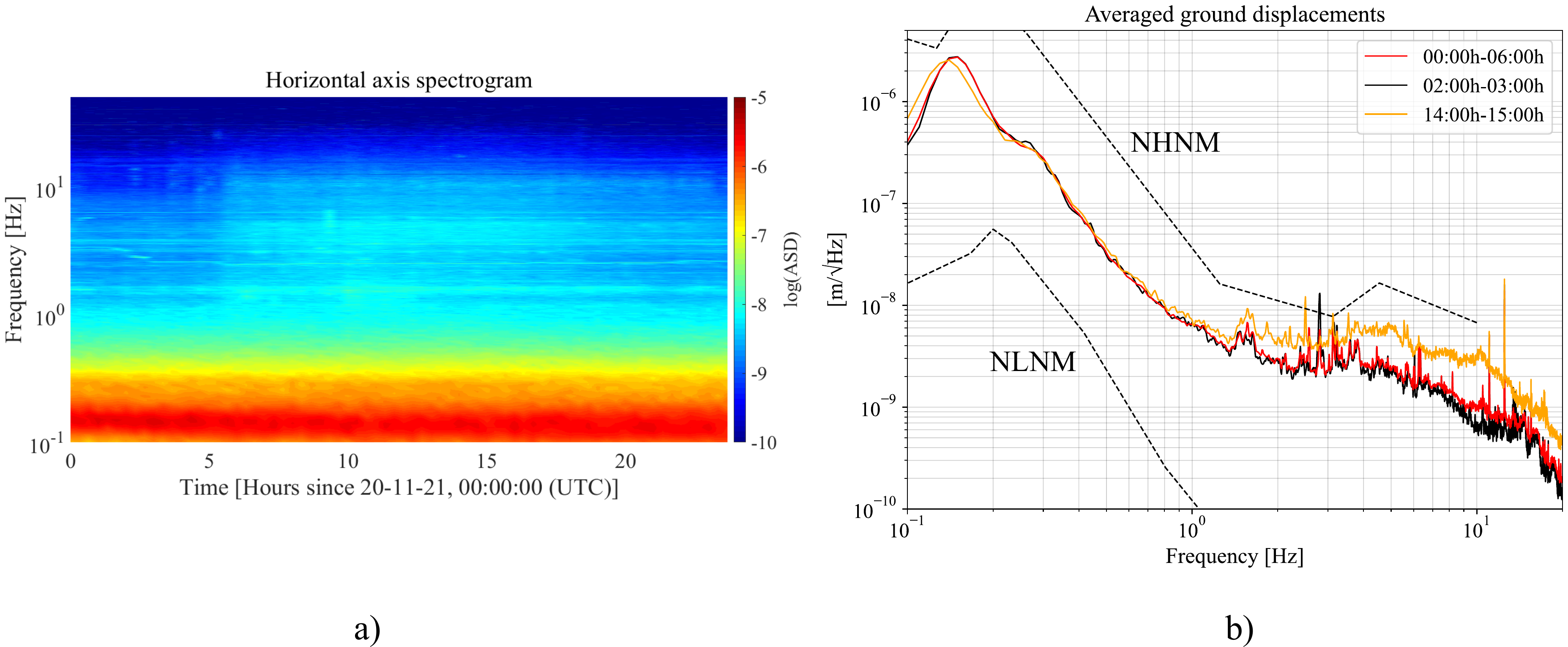}
    \caption{The variability of the ambient field displacement during a 24 hours period. The spectrogram in Figure a) shows the continuous variation in the magnitude of the ground displacement for one horizontal axis with the colorbar units according to the exponents of the amplitude spectral density. Plot b) shows the displacement amplitude $S_{H}$ of the ground during three intervals of the day. The red curve corresponds to the data used for the estimation of seismic noise component. The curves are compared to the low noise model (NLNM) and new high noise model (NHNM) from Peterson \cite{Peterson}. The corresponding spectrogram of the vertical ground displacement shows a variability similar to the horizontal one.}
    \label{fig:spectrogram}
\end{figure}

The measurement of the ground displacement was taken on the laboratory concrete slab using a triaxial Trillium T240 seismometer. The seismic variability during a 24 hours measurement duration can be seen is the spectrogram from Figure~\ref{fig:spectrogram} as well as the variation in the ground displacement measured at different times during the day. Being in an inner-city location, a large amplitude contribution from anthropogenic noise is clearly visible between 1 to \SI{20}{Hz}, with slight reduction during night-time. The output velocity signal of the Trillium T240 is integrated and the mean density per frequency bin computes the triaxial ground displacement set: $S_{V}(f)$ for the vertical axis and $S_{E}(f)$, $S_{N}(f)$ for the eastern and northern axes respectively. A cumulative horizontal displacement spectrum is given by $S_{H} = \sqrt{S_{E}(f)^{2} + S_{N}(f)^{2}}$.

\subsubsection{Suspension transfer function and noise propagation}
\label{sec:sus_transfer}
\begin{figure}[htbp]
    \centering
    \includegraphics[width=.54\linewidth]{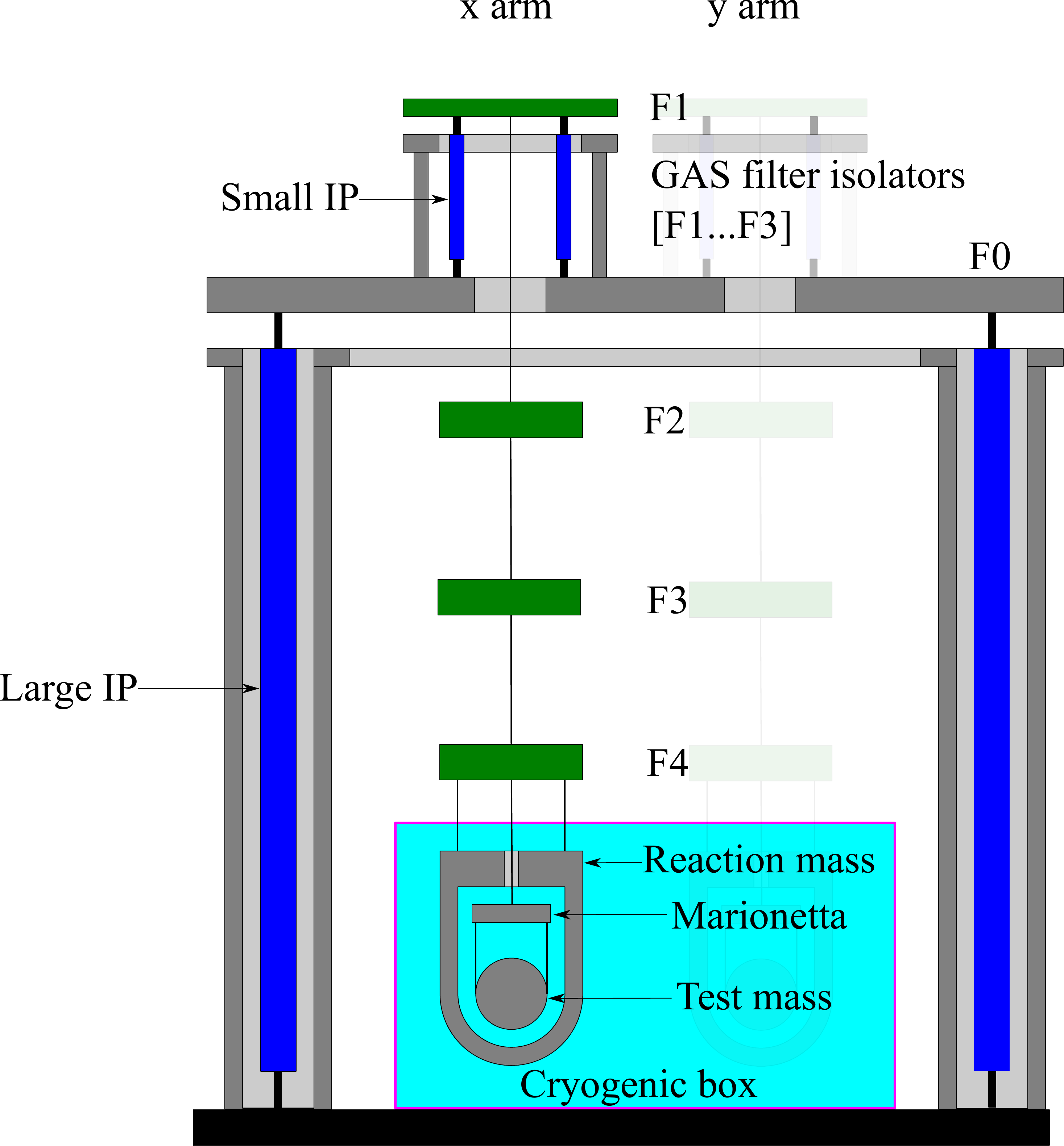}
    \caption{Simplified representation of the ETpathfinder suspension system. The blue rectangles show the two stages of inverted pendulums (four large IP stages for the entire suspension structure and three small IP stages per each suspension chain) which together with five cascaded bodies provide an attenuation that falls with $1/f^{14}$ after the pendulum mode of the test mass. The green rectangles provide vertical attenuation via geometric anti-spring systems. The cryogenic payload comprises of the reaction mass, the marionetta and the test mass.}
    \label{fig:suspension}
\end{figure}

In order to shield the core optics from the seismic propagation and to achieve the sensitivity target of $10^{-18}$ m/$\sqrt{\text{Hz}}$ at \SI{10}{Hz}, multi-stage vibration isolators will be used. The horizontal seismic motion, which acts in the measurement direction and parallel to the leaser beam, needs to be attenuated by at least nine orders of magnitude. The suspension system typically translates around 0.1\% of vertical motion into horizontal displacement, therefore the vertical motion needs to be attenuated by at least six orders of magnitude~\cite{ETpf2}. The ETpathfinder isolation system resembles the mature technology of the Virgo superattenuators~\cite{sta}. A simplified schematic of the isolation system is shown in Figure~\ref{fig:suspension}. Its horizontal isolation is achieved by seven stages of pendulums and the vertical isolation by four stages containing cantilever springs. The first two stages are inverted pendulums pre-isolators in three degrees of freedom (longitudinal, yaw and vertical), tuned for low-frequency. A simple approach to estimate the vibration transmissibility through the suspension is to approximate it as a linear time-invariant system (LTI). The transfer function of the mechanical system is constructed from the \textit{state-space} representation of the LTI model. The calculation is performed for two degrees of freedom: the longitudinal motion and the vertical motion with the 0.1\% coupling factor. The following simplifications are made:
\begin{itemize}
    \item The suspension wires are assumed to be massless, thus the violin modes are not accounted for.
    \item The suspension wires connect to the rigid bodies at their center-of-mass. This simplifies the dynamics of the system and removes any effects from the rotational degrees of freedom.
    \item The inverted pendulum stages are modelled as rigid mechanical rods with their center-of-mass at medium point.
\end{itemize}

\begin{figure}[ht]
     \centering
     \begin{subfigure}[b]{0.32\textwidth}
         \centering
         \includegraphics[width=\textwidth]{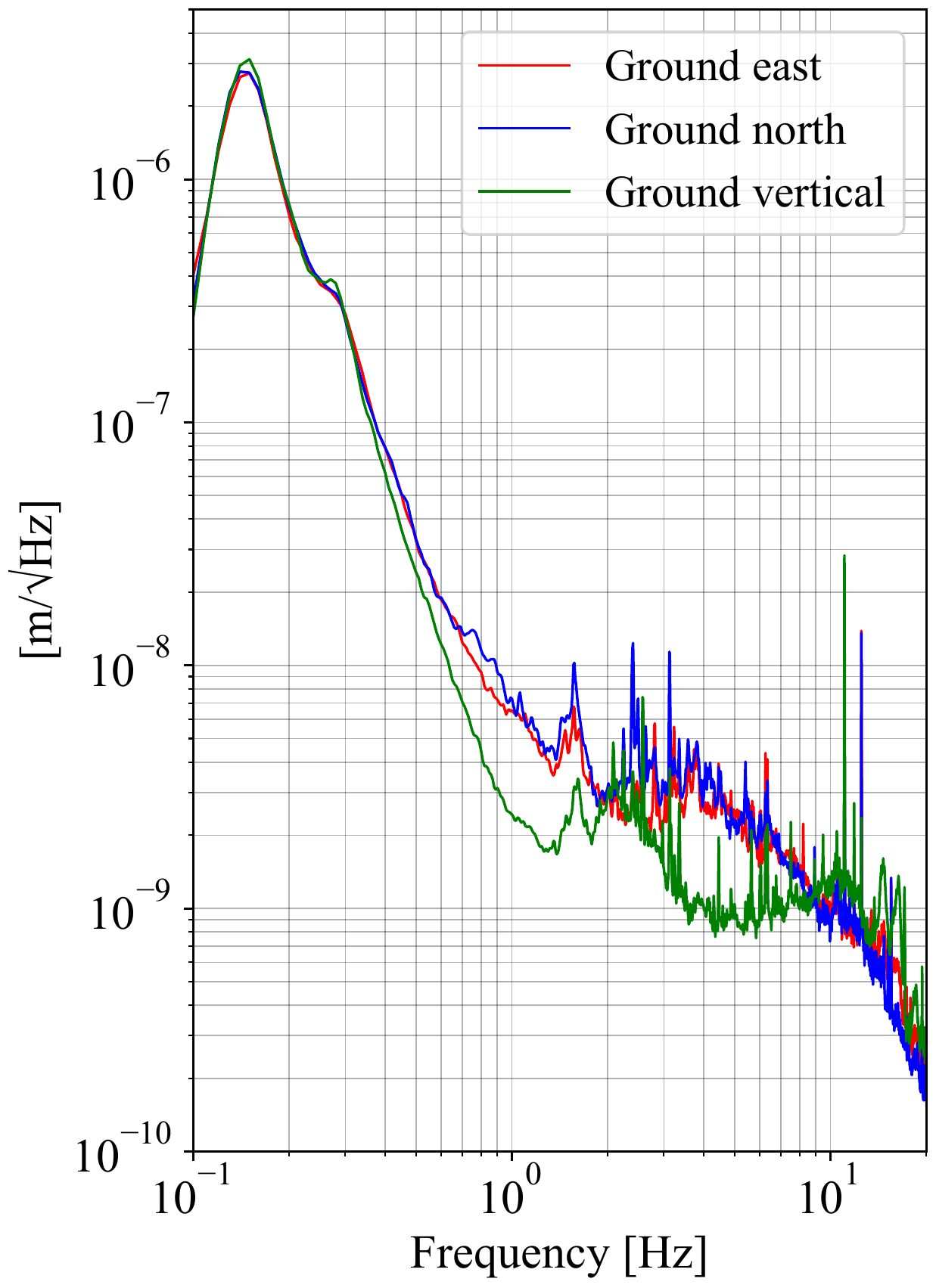}

     \end{subfigure}
     \hfill
     \begin{subfigure}[b]{0.32\textwidth}
         \centering
         \includegraphics[width=\textwidth]{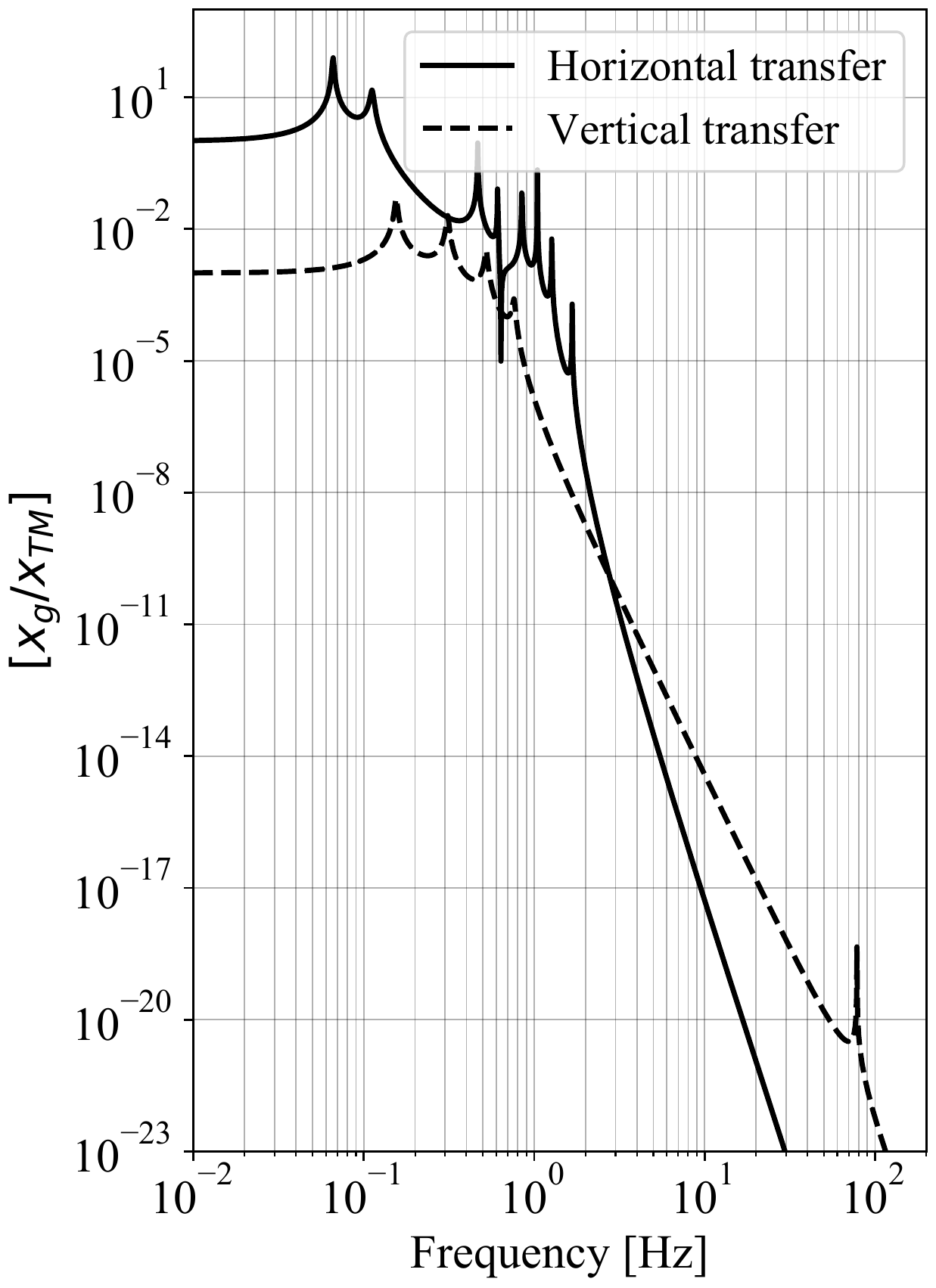}

     \end{subfigure}
     \hfill
     \begin{subfigure}[b]{0.32\textwidth}
         \centering
         \includegraphics[width=\textwidth]{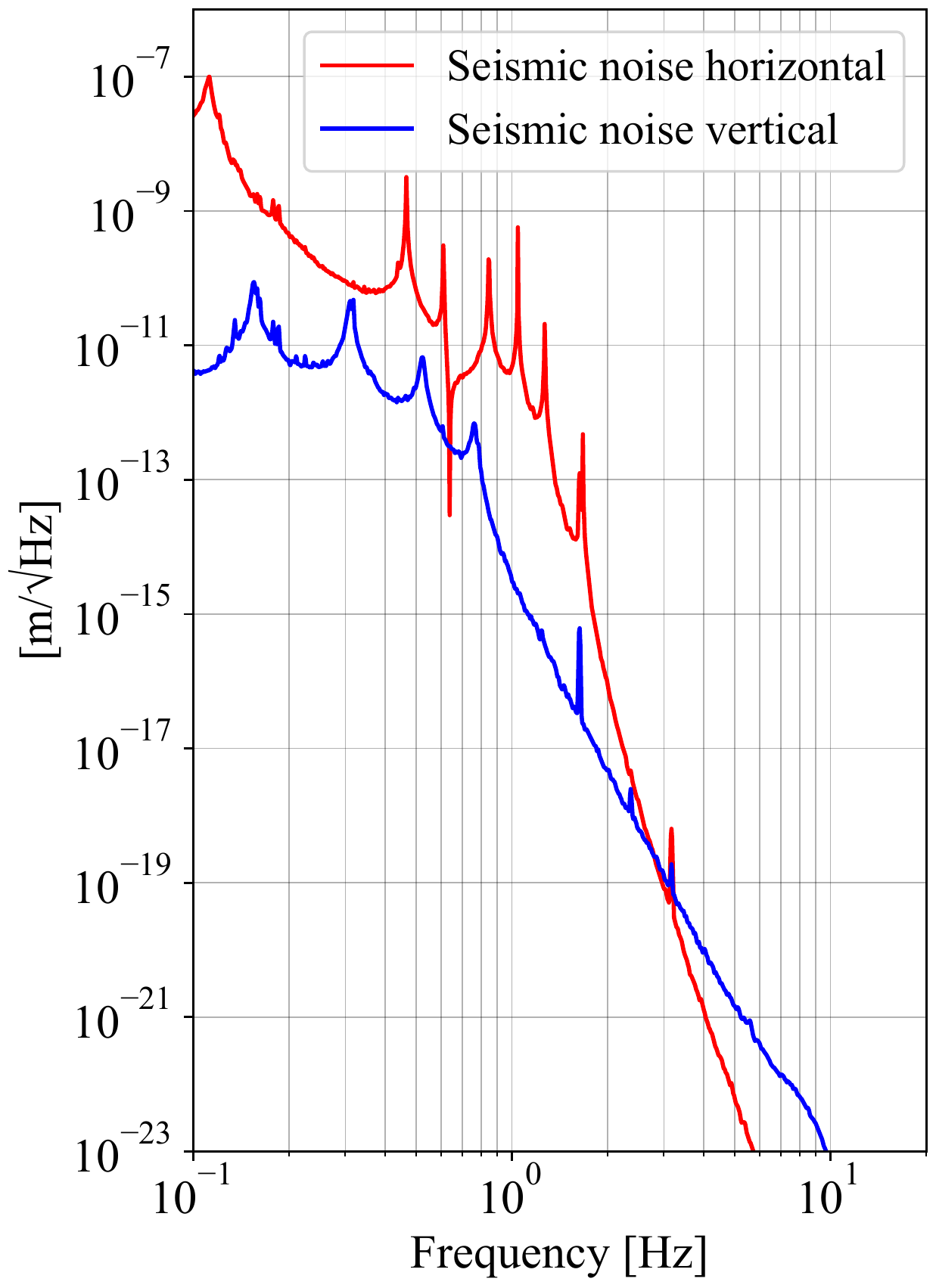}

     \end{subfigure}
        \caption{The estimation of seismic noise. The left plot shows the amplitude spectral density of the ground motion in three orthogonal directions after 6 hours of integrated data, as shown in Figure~\ref{fig:spectrogram}. These curves are multiplied by the suspension transmissibility in the middle figure. The vertical transfer is created assuming a coupling of 0.1\% to the longitudinal degree of freedom. The resultant residual seismic noise of the test masses is shown in the right plot.}
        \label{fig:seismic_plots}
\end{figure}

The transfer function together with the projected seismic noise is presented in Figure~\ref{fig:seismic_plots}.
\subsection{Newtonian noise}

The isolated test mass is susceptible to gravitational forces by either static or travelling mass-density oscillations. This noise source is called Newtonian noise or gravity-gradient noise. It was recognized since the beginning of the field, with some important work being developed early on~\cite{SaulsonNN, Raiss}. Here we used the formalism from~\cite{Harms} which is based on the previous work showing that the dominant contribution to Newtonian noise comes from the seismic surface Rayleigh waves~\cite{Ht,Bc}. These waves produce correlated displacement fluctuations below the test mass.

Here the gravity gradient noise is estimated using the ambient seismic measurement in the vertical axis, $S_{v}(f)$, shown in Section~\ref{sec:sus_transfer}. An important characteristic of ETpathfinder is its short armlengths and the folded interferometer topology which locates analogous test masses close to each other. In this case the seismic wavelength is larger that the distances between adjacent test masses. Therefore, the test masses move in phase when they respond to passing Rayleigh waves and this strongly suppresses the gravity-gradient effect from the differential measurement of the interferometer.

The power spectral density of differential acceleration propagating along the $x$ axis, along the armlength $L$, $S_{nn}(f)$ is estimated as~\cite{Harms}:

\begin{equation}
\label{eq:NN}
    S_{nn}(f) = (2\pi G \rho_{0} e^{-hk_{\rho}} \gamma(\nu))^{2} \times \frac{1}{2} \times S_{v}(f) \times \begin{bmatrix} 1 - 2J_{0} (k_{\rho}L) + 2J_{1} (k_{\rho}L)/(\rho L) \\ 1-2 J_{1} (k_{\rho}L)/(\rho L) \\ 2-2 J_{0} (k_{\rho}L)  \end{bmatrix}.
\end{equation}
Here $G$ is the gravitational constant, $\rho_{0}$ is the density of the ground which is assumed to be 1800 $\text{kg}/\text{m}^{3}$. The height above ground $h$ is \SI{1.2}{m}. The term $k_{\rho} = \frac{2 \pi f}{c}$ requires knowledge about the Rayleigh wave speed $c$. A generally good approximation of $c$ is 200 m/s, similar to the measurements performed at the LIGO Livingston site~\cite{Harms2}. The term $\gamma(\nu)$ is related to the dispersive properties of the site's ground and it is set to 0.8 here. The equations in the bracket correspond to three directional averaged response terms where $J_{n}(x)$ is a Bessel function of order $n$.

Under the assumption that the seismic field is isotropic, each acceleration component of Equation~\ref{eq:NN} becomes independent at a point measurement. This makes it possible to assume that two orthogonal correlations can be introduced by multiplying the two components, in the $x$ and $y$ directions, with a simple estimate of the Newtonian noise assuming no correlations. Thus, the $x$ term in Equation~\ref{eq:NN} assumes an effective length $L$ of \SI{9.2}{m} and the $y$ term assumes an effective length of \SI{0.45}{m}, which is the distance between the two center of masses of adjacent input or end mirrors.

\begin{figure}[htbp]
    \centering
    \includegraphics[width=.7\linewidth]{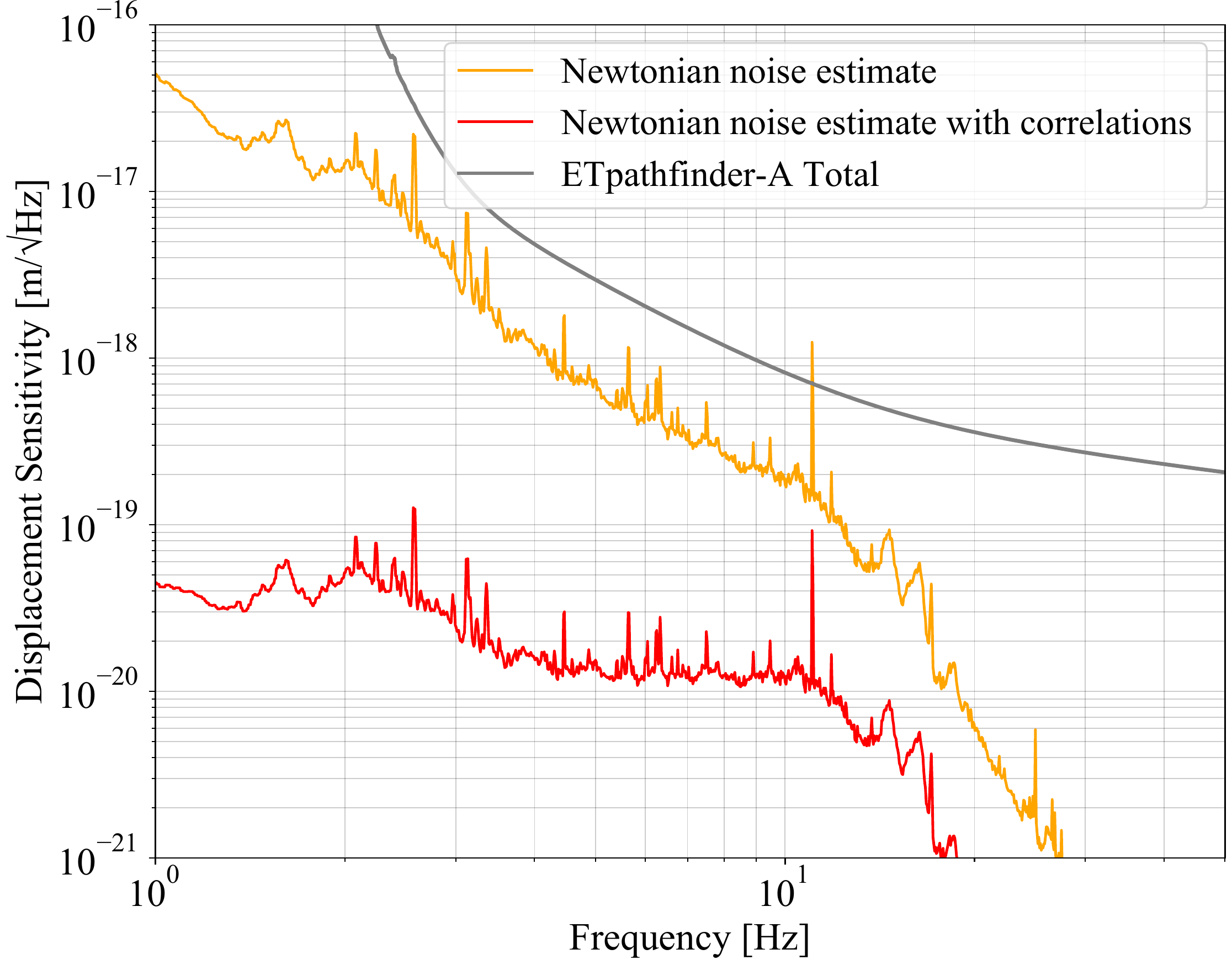}
    \caption{The amplitude spectral density of the Newtonian noise. By adding the simple seismic correlations, both across the armlength and orthogonal to the adjacent test masses, a substantial suppression of noise is obtained.}
    \label{fig:newtonian}
\end{figure}

The estimation of the displacement spectral density is then calculated as $\sqrt{S_{nn}(f) \times 2}/(2 \pi f)^{2}$, for a double correlated effect across the full interferometer. The results are shown in Figure~\ref{fig:newtonian}. Compared to the estimation of the gravity gradient noise for singular test masses, the estimation using orthogonal correlations is strongly suppressed. However, depending on the measurement duration, the variation in seismic activity and possible anisotropies in the propagating field, the resulting Newtonian noise will most probably lie between the two estimates shown. This will not affect the target displacement sensitivity at lower frequencies.

\subsection{Suspension thermal noise}
The elements in the cascaded suspension structure fluctuate thermally as a result of mechanical dissipation, viscous damping or inelastic behaviour. Suspension thermal noise is expressed via the FD theorem. In order to perform the calculation, two things have to be defined. One is the test mass admittance and the other one is the departure from the elastic oscillatory regime~\cite{s1}. The admittance is the inverse of the mechanical impedance defined in the FD theorem by Equation~\ref{eq:FDT}:

\begin{equation}
Y(f) = Z(f)^{-1} = \frac{v(f)}{F_{thermal}} = \frac{2\pi i f x(f)}{F_{thermal}}.
\end{equation}
Here $F_{thermal}$ corresponds to the thermal driving force and $v(f)$ is the resulting velocity, derivative of the readout variable $x(f)$. For a suspended pendulum, the driving force producing the harmonic oscillation can be understood as a Langevin stochastic force on the body. Thus, the force response $\frac{x(f)}{F_{thermal}}$ is estimated by the transfer function formalism presented in the Section~\ref{sec:sus_transfer}. The present estimation takes into account not only the horizontal force response, but also the vertical response which creates a dominant factor in the noise.
This is allowed, since the impedance formulation from the FD theorem characterises a system for a superposition of eigenstates~\cite{PhysRev.83.34}. However, for a more complete description of this noise, all six degrees of freedom of the test mass have to be considered~\cite{Gonzalez2000SuspensionsTN}.

The anelasticity is the system characteristic that causes dissipation of energy. This is included here in the trivial way by expressing the stiffness of the suspension wire as a complex element in the frequency domain:
\begin{equation}
k \to k_{spring}(1+i\phi).
\end{equation}
Here $\phi$ expresses the phase angle in radians for which the response $x(f)$ lags behind the driving force $F_{thermal}$. It is the structural loss angle encountered in the definition of thermal noises described earlier and known to be constant with frequency~\cite{s1,s2}.

\begin{figure}[htbp]
    \centering
    \includegraphics[width=.7\linewidth]{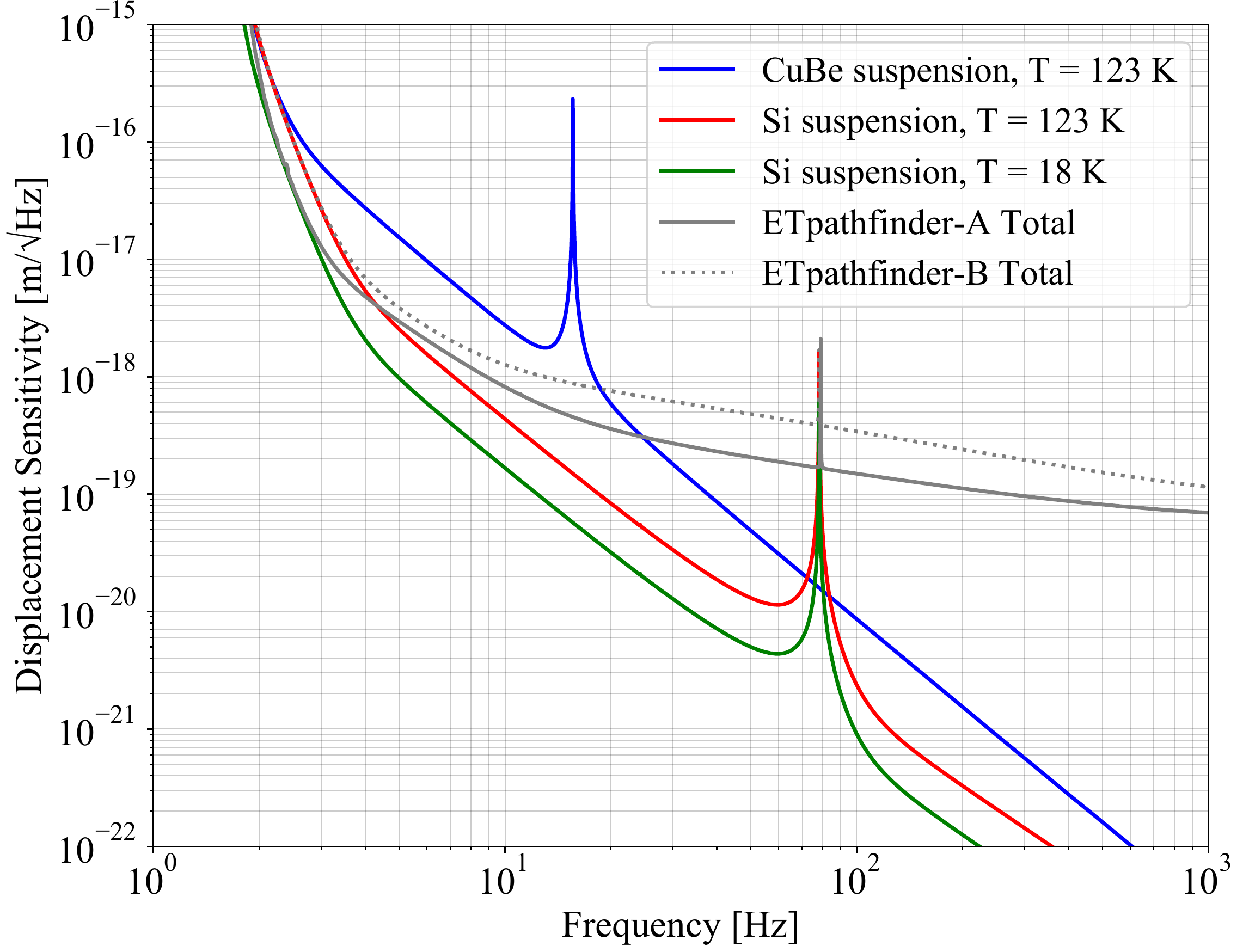}
    \caption{Amplitude spectral density of the suspension thermal noise for different last stage suspension wires and temperatures. The two resonances at \SI{17}{Hz} and \SI{79}{Hz} correspond to the vertical bounce mode of the mirror for each suspension fibre diameter. With lower mechanical dissipation, silicon suspensions provide a five-fold broadband improvement in the suspension thermal noise. Reducing the temperature from \SI{123}{K} to \SI{18}{K}, there is a decrease in the noise by a factor $\sqrt{\frac{123}{18}}$.}
    \label{fig:STN}
\end{figure}

In the initial phase of the ETpathfinder, the silicon test masses will be suspended by Copper Beryllium (CuBe) fibres which are suitable for their low surface thermoelastic dissipation at 123 K. Their diameter size is designed for a tensile strength about a third of the yield strength of the material which is about \SI{1.2}{GPa}. Additionally, in the later stages of the project, ETpathfinder will integrate silicon suspension fibres or ribbons. Their design is driven by the material strength and the thermal conductive power required. We consider \SI{80}{MPa} to be the minimum yield strength of silicon suspension fibres \cite{Birney}. In Figure~\ref{fig:STN}, the baseline thermal noise curve is shown, together with silicon suspension estimations at two different temperatures. The relevant parameters are shown in Table~\ref{tab:STN}.

\begin{table}[htbp]
\caption{Suspension fibres characteristic parameters. All the values correspond to \SI{123}{K} temperature. For the silicon fibres at \SI{123}{K}, the values are assumed the same as for the \SI{18}{K} ones, designed for conductive cooling capability.}

\centering
\footnotesize
\begin{tabular}{@{}l*{15}{l}}
\br
Parameter&Material&Loss angle& Length [m]& Fibre diameter [m]\\
\mr
Marionetta wire& Ti grade 5& $10^{-5}$& 0.6&$8\times 10^{-4}$\\
Test mass wire 1& CuBe& $10^{-5}$& 0.4&$150\times 10^{-6}$\\
Test mass wire 2& Si& $10^{-9}$& 0.4&$7\times 10^{-4}$\\
\br
\end{tabular}
\label{tab:STN}
\end{table}

\subsection{Residual gas noises}
The residual gas noise sources are given by the combination of the arm-length gas scattering noise and the test mass gas damping noise. The statistical fluctuations in the gas column density across the interferometer beam tube produce molecular interaction with the laser field which give rise to optical path change that can mask the desired readout signal~\cite{l49}. The corresponding power spectral density takes the form~\cite{ResGas1}:

\begin{equation}
    S_{L}(f)=\frac{4 \rho(2 \pi \alpha)^{2}}{v_{0}} \int_{0}^{L} \frac{\exp \left[-2 \pi f w(z) / v_{0}\right]}{w(z)} d z.
\end{equation}
Here, $\rho$ is the gas number density for each molecule, $\alpha$ is the polarizability, $v_{0} = \sqrt{\frac{2k_{b}T}{m}}$ is the molecules most probable velocity, $L$ is the armlength and $w(z)$ is the Gaussian beam radius.

\begin{figure}[htbp]
    \centering
    \includegraphics[width=.7\linewidth]{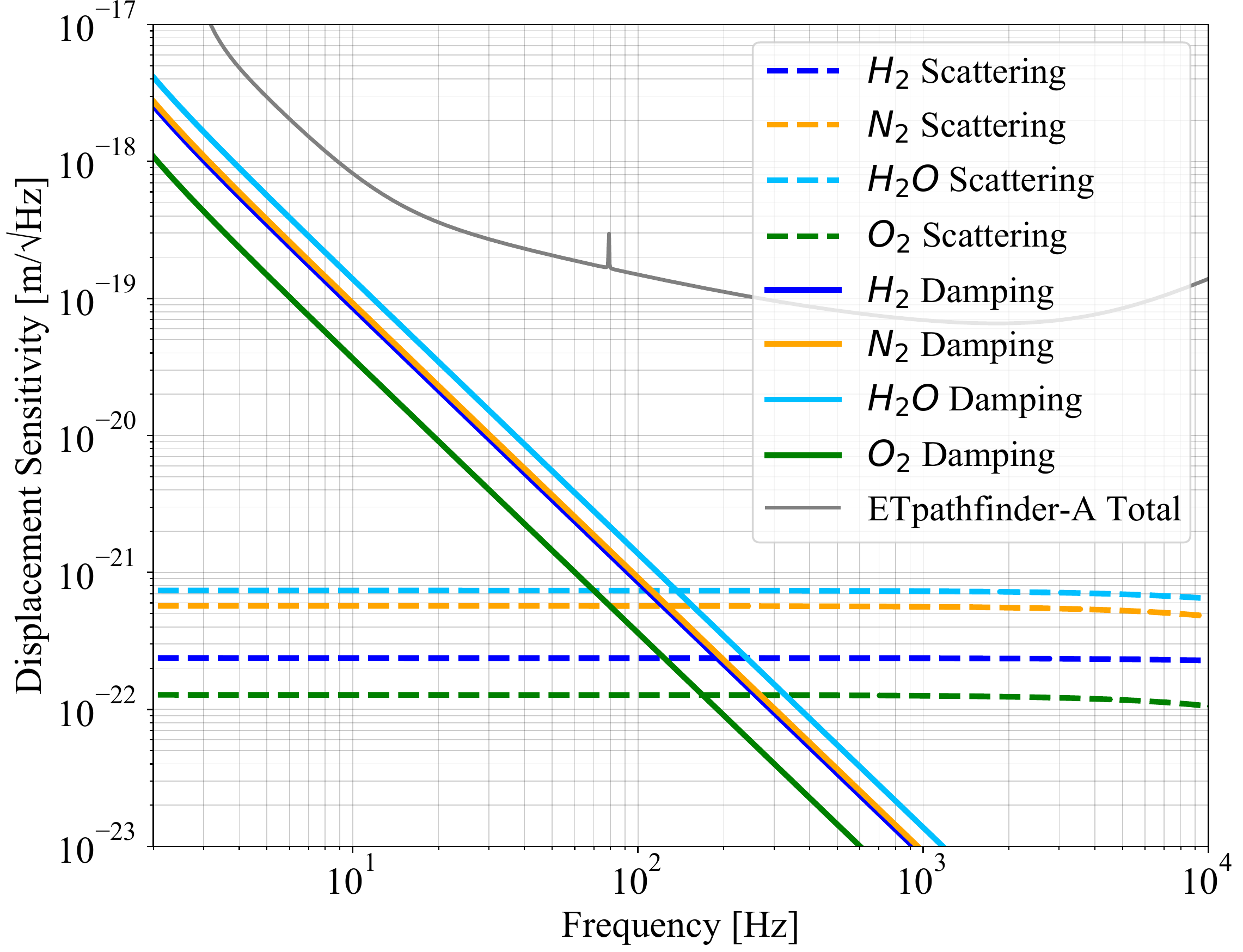}
    \caption{Displacement spectral densities of the gas damping noise and the gas scattering noise for different molecular species. The combination of high polarizability and partial vacuum pressures makes $\text{H}_{2}\text{O}$ excess gas noises spectrum dominate. Even tough the small beam radius is strongly affected by residual scattering, the noise effect is integrated over a very short armlength and does not constitute a limiting factor.}
    \label{fig:res_gas}
\end{figure}

At the end of the arm-length, the free test mass exchanges momentum with the residual particles which results in a transitional damping coefficient $\beta_{tr}$, proportional to the gas column pressure $p$ \cite{CAVALLERI20103365}. The FD theorem then relates the damping coefficient with the corresponding power spectrum of the Brownian noise $x(f) = 4k_{b}T\beta_{tr}$. Even tough the proximity-enhanced gas damping can also be significant~\cite{PhysRevD.84.063007}, it is not considered in our estimation. Figure~\ref{fig:res_gas} shows the two residual gas effects for different molecular species. The associated pressures for each molecule are taken from the acceptance tests of one of the vacuum tube sections and the first delivered vacuum tower. These are presented in Table~\ref{tab:res_gas}.

\begin{table}[htbp]
\caption{Molecular species characteristic pressures extracted from the the residual gas analyser acceptance tests of the first delivered vacuum tower and the first beam tube section.}
\footnotesize
\centering
\begin{tabular}{@{}lll}
\br
Molecule & Bench tower pressure [Pa] & Beam tube pressure [Pa]\\
\mr
$\text{H}_{2}$ & $1.2 \times 10^{-6}$ &$3.8 \times 10^{-7}$\\
$\text{N}_{2}$ &$4.0 \times 10^{-7}$ &$1.2 \times 10^{-7}$\\
$\text{H}_{2}\text{O}$ &$1.1 \times 10^{-6}$ &$3.3 \times 10^{-7}$\\
$\text{O}_{2}$ &$5.8 \times 10^{-8}$ &$7.0 \times 10^{-9}$\\
\br
\end{tabular}
\label{tab:res_gas}
\end{table}




\section{Conclusions}
\label{sec:Conclusions}
ETpathfinder is a cryogenic prototype interferometer that provides a unique test environment for the building blocks that are required for the future gravitational-wave detectors such as the Einstein Telescope and Cosmic Explorer.

In the first running phase at \SI{123}{K}, ETpathfinder-Light, the displacement sensitivity of the interferometer will be limited by the suspension thermal noise in low frequencies, the coating thermal noise in middle band and the shot noise at very high frequencies. At this temperature, the thermo-optic coefficient for crystalline silicon becomes relevant in relation to the very small beam diameters on the optics. The ITM thermorefractive noise is thus reduced by increasing the arm-cavity finesse value to 2050. This allows the middle sensitivity band to be limited by the coating Brownian noise which provides a testing ground directly linked to the limiting noise of the 3rd generation detectors even at a transitory temperature of operation.

At cryogenic temperatures below \SI{20}{K}, the \SI{1550}{nm} interferometer reaches the low-noise target sensitivity of $\SI{1e-18}{m\per\sqrt{Hz}}$ at \SI{10}{Hz} from the substantial noise improvement at low frequency by the integration of silicon suspension fibres. The reduction in suspension thermal noise also relaxes the sensitivity of the 2 microns interferometer. Furthermore, the low substrate thermal noise at \SI{18}{K} temperature can be achieved due to vanishingly small thermo-elastic coefficient. To accurately project this noise source at very low temperatures and for smaller beam diameters on the test masses, the full series solution of the power spectrum estimation is used. The reduction leaves a big gap with the limiting coating thermal noise.

The coating Brownian noise-limited sensitivity gives the opportunity for testing different coating strategies. An interesting example of this is the implementation of amorphous silicon nitride as a low refractive-index material in combination with amorphous silicon. The resulting coating Brownian noise will be substantially reduced. As a result, the total displacement sensitivity would fundamentally be limited only by the well understood quantum-noise over a wide frequency band, and therefore allows to investigate and tackle a variety of technical noise sources linked to the operation at cryogenic temperatures. The degree to which these noise sources are attenuated and the means of achieving that are directly relevant to the technical design of the 3G detectors.

In a latter stage of operation and depending on the outcome of the initial running phases, ETpathfinder will operate at a single temperature and laser wavelength. The test masses of this experimental phase will resemble the ET scale and will be integrated in single cryostats. This phase is planned to run at the end of the decade in order to directly provide useful inputs to ET.

\section{Acknowledgements}
\label{sec:funds}
The full data and simulation developed for this study were integrated into pyGWINC and are available here: \url{https://git.ligo.org/et-pathfinder/noise-budget/-/releases/sensitivity-paper-2022}. The authors would like to thank Jan Harms and Maria Bader for very useful discussions regarding the analytical and the experimental aspects of the Newtonian and seismic noises.
\subsection{Funding}
The ETpathfinder project in Maastricht is funded by Interreg Vlaanderen-Nederland, the province of Dutch Limburg, the province of Antwerp, the Flamish Government, the province of North Brabant, the Smart Hub Flemish Brabant, the Dutch Ministry of Economic Affairs, the Dutch Ministry of Education, Culture and Science, and by own funding of the involved partners.

\appendix
\section{Collection of the most important parameters}
\label{sec:AppendixA}
\begin{table}[htbp]
\caption{Summary of the most important parameters for both ETpathfinder interferometers at two different temperatures. Both interferometers have a similar baseline of \SI{9.2}{m} and their seismic isolation system is similar with last stage suspension fibres length of \SI{0.4}{m}.}

\centering
\footnotesize
\begin{tabular}{@{}l*{80}{l}}
\br
Parameter&ETpathfinder-Light&ETpathfinder-A&ETpathfinder-B\\
\mr
Temperature [K] & 123& 18 & 123\\
Wavelength [nm] &1550 & 1550 &2090 \\
Arm-cavity finesse &2050 & 2050 &2050 \\
Test mass weight [kg] & 3.2 & 3.2 & 3.2 \\

Beam waist [m] & $1.8 \times 10^{-3}$ & $1.8 \times 10^{-3} $ & $2.12 \times 10^{-3}$ \\
Beam radius at test mass [m] & $2.2 \times 10^{-3}$ & $2.2 \times 10^{-3} $ & $2.56 \times 10^{-3}$ \\
Substrate young modulus [Pa]& $155.8\times 10^{9}$ & $162.0\times 10^{9}$ &$155.8\times 10^{9}$
\\
Substrate thermal conductivity [W/(m$\cdot$K)] & 700& 3000& 700 
\\
Thermal expansion coefficient [1/K] & $1\times 10^{-9}$ & $1\times 10^{-9}$  & $1\times 10^{-9}$
\\
Substrate specific heat [J/(kg$\cdot$K)] & 333 & 3.5 & 333
\\
Thermorefractive coefficient & $1\times 10^{-4}$ & $1.1\times 10^{-6}$ & $1\times 10^{-4}$
\\
Substrate loss angle &$1.25\times 10^{-9}$ & $1.25\times 10^{-9}$ & $1.25\times 10^{-9}$
\\

Last stage suspension material & Copper Beryllium & Silicon & Silicon \\
Last stage suspension fibres diameter [m] & $1.5 \times 10^{-4}$ & $7 \times 10^{-4}$ & $7 \times10^{-4}$ \\
Coating $\phi_{high}$ & $5.7\times 10^{-4}$ &$5.6\times 10^{-4}$ & $5.7\times 10^{-4}$ \\
Coating $\phi_{low}$ & $4.8\times 10^{-4}$ &$9.2\times 10^{-4}$ & $4.8\times 10^{-4}$  \\
\br
\end{tabular}
\label{tab:parameters}
\end{table}

\newpage
\section{Intermediate step, ETpathfinder-A at 123 K}
\label{sec:AppendixB}
\begin{figure}[htbp]
    \centering
    \includegraphics[width=.7\linewidth]{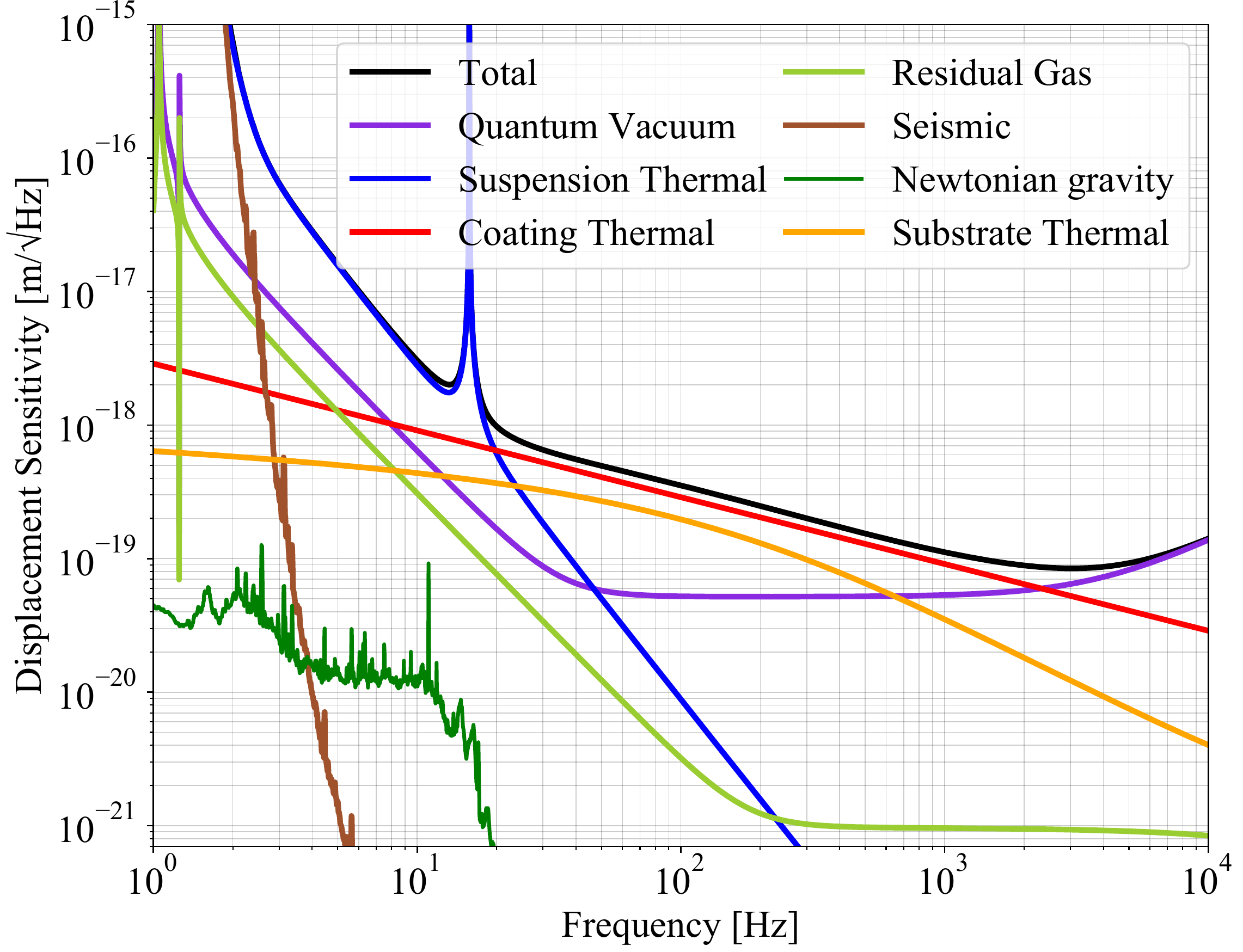}
    \caption{Projection of displacement sensitivity for the initial phase of ETpathfinder.}
    \label{fig:ETpfA_light}
\end{figure}

\newpage
\section*{References}
\bibliographystyle{iopart-num}
\bibliography{bib_file}

\end{document}